\documentclass[prd,twocolumn,aps,reprint,nofootinbib,showpacs,superscriptaddress]{revtex4-1}

\usepackage[bookmarks,pdfhighlight=/O,pdfstartview=FitH]{hyperref}
\usepackage{url}

\usepackage{graphicx}

\newcommand{\onepion}{single-pion~}
\newcommand{\onenucleon}{single-nucleon~}

\newcommand{\dd}{\mathrm{d}}
\newcommand{\cm}{\; \mathrm{cm}}
\newcommand{\GeV}{\; \mathrm{GeV}}
\newcommand{\MeV}{\; \mathrm{MeV}}

\begin{document}

\title{Neutrino and antineutrino induced reactions with nuclei between 1 and 50 GeV}

\author{O. Lalakulich}
\affiliation{Institut f\"ur Theoretische Physik, Universit\"at Giessen, D-35392 Giessen, Germany}
\author{K. Gallmeister}
\affiliation{Institut f\"ur Theoretische Physik, Johann Wolfgang Goethe-Universit\"at Frankfurt, D-60438 Frankfurt-am-Main, Germany}
\author{U. Mosel}
\email{mosel@physik.uni-giessen.de}
\affiliation{Institut f\"ur Theoretische Physik, Universit\"at Giessen, Germany}

\begin{abstract}
\begin{description}
\item[Background] Nuclear effects can have a significant impact on neutrino-nucleus interactions.
In particular, data from neutrino experiments with broad energy distributions require complex
theoretical models that are able to take all the relevant channels into account
as well as incorporate nuclear effects in both initial and final-state interactions.
\item[Purpose] We investigate neutrino and antineutrino scattering on iron and carbon in the energy range
from 1 to 50 GeV, which is relevant to current and coming experiments (MINOS, NO$\nu$A, and Miner$\nu$a).
\item[Method] The Giessen Boltzmann--Uehling--Uhlenbeck (GiBUU) model, which implements all reaction channels relevant
for neutrino energies under consideration, is used for an investigation of neutrino-nucleus reactions.
\item[Results] Our calculations are compared with the recent NOMAD and MINOS data for the integrated inclusive cross sections.
Predictions are made for the differential cross sections for semi-inclusive final states (pions, kaons, and nucleons) for the MINOS and NO$\nu$A
fluxes.
\item[Conclusions] Nuclear effects in the initial-state interactions may slightly change the inclusive nuclear cross section
as compared to the free nucleon ones. Final-state interactions noticeably change the spectra of the outgoing hadrons.
In the Miner$\nu$a and NO$\nu$A experiments these effects should be visible in the
kinetic energy distributions of the final pions, kaons and nucleons. Secondary interactions play an important role for strangeness production. They make
it very difficult to extract the neutrino-induced strangeness production cross section in experiments using nuclear targets.
\end{description}
\end{abstract}


\maketitle

\section{Introduction}

Neutrino and antineutrino scattering on nuclei for neutrino energies above $30 \GeV$ was studied
in several experiments starting in the 1980s (see Refs.~\cite{Conrad:1997ne,Tzanov:2009zz} for reviews).
Based on muon detection, integrated cross sections were measured with overall precision of 2\%;
they grow linearly with energy,  which agrees with the predictions of the quark parton model.
Double differential cross sections with respect to muon variables were also
measured, allowing one to extract the nucleon structure functions.

Theoretically neutrino reactions at lower energies (a few GeV to a few 10 GeV)  still present a challenge because here
the methods of perturbative QCD are not reliable and because contributions from quasi-elastic (QE) scattering, resonance (RES) production, 
background processes, and deep inelastic scattering (DIS) overlap.
This is especially important in view of the broad energy distributions of neutrino fluxes.
The difficulty lies in choosing an appropriate way to describe the onset  of the DIS processes,
the so-called shallow inelastic scattering (SIS). This requires complex approaches that take all of the relevant channels into account.
The lack of data at intermediate energies has long been an obstacle for a serious test of such approaches.

New and coming experimental results are changing the situation.
Recently, the NOMAD collaboration performed measurements on an a composite target and reported the inclusive neutrino
cross section on an isoscalar target for $E_\nu>4.6 \GeV$ with an accuracy of at least $4\%$ \cite{:2007rv}.
The MINOS experiment reported both neutrino (for $E_\nu>3.48\GeV$)
and antineutrino (for $E_{\bar\nu}>6.07\GeV$) cross sections also on an iron target
with a comparable precision \cite{Adamson:2009ju}.
Cross sections at lower energies on (mostly)  a carbon target
will be measured by the NO$\nu$A experiment.
The Miner$\nu$a experiment intends to perform measurements on
plastic (CH), iron, lead, carbon, water and liquid helium targets,
which would directly  allow to compare nuclear effects on various nuclei.
In addition to muon detection, this experiment will also be able to resolve various final states
by identifying the tracks of the outgoing hadrons.

On the theory side only very few papers have tried to describe not only the very high, but also the intermediate energy region. One of the most complete ones is the work by Kulagin and Petti \cite{Kulagin:2007ju} (see also Haider et al.\ \cite{Haider:2011qs}), who have presented a detailed description of neutrino inelastic inclusive scattering in terms of nuclear structure functions. In their work these authors also pay attention to a number of nuclear effects, such as Fermi motion and binding, and how these affect the extraction of structure functions. Our work is similar in spirit, but we aim for a practical implementation in an event generator that allows one to calculate not only inclusive cross sections, but also semi-inclusive ones as they will be measured by Miner$\nu$a, for example.

In this paper we study neutrino and antineutrino scattering  on iron and carbon.
Our results  are compared with the recent MINOS and NOMAD data for inclusive cross sections.
Predictions are also made for the spectra of the outgoing hadrons for the MINOS and NO$\nu$A neutrino fluxes.

\section{GiBUU transport model  \label{gibuu}}

The GiBUU model has been developed
as a transport model for nucleon-, nucleus-,  pion-, and electron-induced collisions from
some MeV up to tens of GeV. Several years ago neutrino-induced interactions were
also implemented for the energies up to a about 2 GeV \cite{Leitner:2006ww,Leitner:2006sp}
and, recently, the GiBUU code was extended to describe also the DIS processes for neutrino reactions.

Thus, with GiBUU it is possible to study various elementary reactions
on  nuclei within a unified framework~\cite{Buss:2011mx}. This is particularly important for broad-beam
neutrino experiments, which inherently average over many different reaction mechanisms.
Relevant for the present investigation is also the fact that the method and code have been widely tested for
photon-induced as well as for electron-induced reactions in the
energy regime from a few hundred MeV to 200 GeV \cite{Effenberger:1999jc,Krusche:2004uw,Buss:2007ar,Gallmeister:2007an,Leitner:2008ue}. To be applicable for such a wide energy range the GiBUU model uses relativistic kinematics throughout as well as relativistic nuclear dynamics as expressed, e.g.,  in way how the mean-field potentials and cross sections are treated. Also the transport equation itself is covariant; for further details on these latter points, see Ref.~\cite{Buss:2011mx}.
We stress that the calculations reported in the present paper are done without any fine tuning to the
data discussed here with the default parameters as used in the general GiBUU framework.

\begin{figure}[hbt]
\includegraphics[width=\columnwidth]{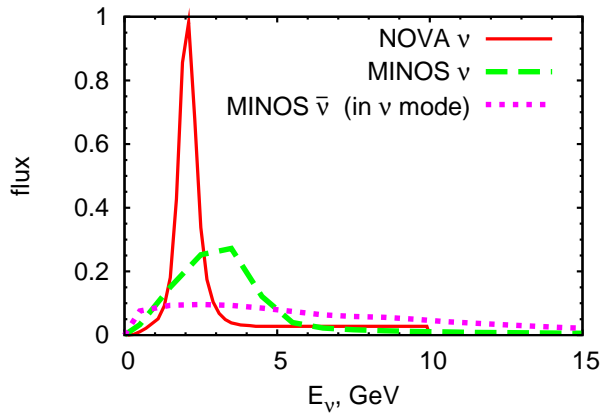}
\caption{(Color online) Energy distributions for NO$\nu$A neutrinos (solid red line) \cite{nova},
MINOS neutrinos (dashed green line),  and MINOS antineutrinos \cite{Bhattacharya:2009zza}
(short-dashed magenta line) normalized to unit area.
The long tails of MINOS neutrino
(up to $50 \GeV$)  and  MINOS antineutrino (up to $38 \GeV$) fluxes are not shown.}
\label{fig:flux}
\end{figure}

As already mentioned, broad band neutrino and antineutrino interactions with nucleons may result in several different channels. This is
evident when looking at the energy distributions (see Fig.~\ref{fig:flux}) of the two experiments, NO$\nu$A and MINOS, that will be discussed later.
The MINOS neutrino and antineutrino fluxes here are those used by the MINOS collaboration \cite{Adamson:2009ju, Bhattacharya:2009zza}
to determine the total cross sections, that is,
the neutrino and antineutrino parts of the low-energy NuMI beam in neutrino mode. The same beam is used by the Miner$\nu$a experiment.
The NO$\nu$A flux is the neutrino part of the 14 mrad off-axis medium-energy NuMI beam in neutrino mode \cite{nova}.

At hadronic invariant masses below the \onepion{} production threshold, $W<1.08 \GeV$, only the QE processes
$\nu n \to \mu^- p$ and $\bar\nu p \to \mu^+ n$ are possible.
At $W>1.08 \GeV$ the \onepion{} production channel opens. The biggest contribution comes from production through
the $\Delta$ [$P_{33}(1232)$] resonance.
With increasing $W$, excitations of resonances with higher masses, followed by their subsequent decays,
give increasing contributions. For the prominent $\Delta$ resonance, the $\Delta\to N \pi$ channel
saturates the width at the $99\%$ level.  For higher resonances, two and more pions,
as well as other mesons, can be produced in the final state.  In the GiBUU code, electroweak
excitations of 13 resonances are accounted for, whose electromagnetic properties are taken from the MAID
analysis \cite{MAID,Drechsel:2007if}, while their axial couplings are assumed to be given by PCAC and dipole form factors.
These processes as well as the  non-resonant background, which gives a noticeable  contribution in the same
$W$ region, are discussed in Ref.~\cite{Leitner:2008ue}.

High-mass resonances overlap: with increasing $W$, the individual resonances become
less and less distinguishable. Thus, at high energies neutrino scattering
needs to be described as a DIS process in terms of quark and gluon degrees of
freedom. Within GiBUU this is done via  the \textsc{pythia} event generator version 6.4.26 \cite{Sjostrand:2006za}.
In it cross sections are obtained using the descriptions for the hard partonic processes,
known from the Standard Model and QCD. Those are convoluted with the quark distribution functions,
for which various known parametrizations can be used; the default is CTEQ 5L, a leading-order fit.
The model separates elastic, single-diffractive, double-diffractive, and non-diffractive event topologies. The determination of final states is done with the help
of the string fragmentation according to the LUND-String model or cluster collapses for low-mass configurations.

Setting for the moment the difficulty aside to describe each of the possible neutrino channels (resonances, background, DIS) separately, there appears a problem of possible double counting. In the SIS region, i.e.,\  the region in between the resonances and DIS, the same physical events can be considered as originating from decays of high-mass baryonic resonances or from DIS. In this region we switch smoothly from the resonance picture to a DIS description.
While the resonance and \onepion background contributions are smoothly switched off above $W_{\rm RES-1}$ the DIS contribution is switched on above $W_{\rm{DIS-1}}$;
\begin{equation}
\begin{array}{l}
\sigma_{\rm RES}=\sigma_{\rm RES}(W)\frac{W_{\rm RES-2}-W}{W_{\rm RES-2}-W_{\rm RES-1}}
\\
\mbox{for   }  W_{\rm RES-1}<W<W_{\rm RES-2} ,
\\[4mm]
\sigma_{\rm DIS}=\sigma_{\rm DIS}(W)\frac{W-W_{\rm DIS-1}}{W_{\rm DIS-2}-W_{\rm DIS-1}}
\\
\mbox{for   } W_{\rm DIS-1}<W<W_{\rm DIS-2}
\end{array}
\label{Wcuts}
\end{equation}
The default parameters are $W_{\rm RES-1}=2.0\GeV$, $W_{\rm RES-2}=2.05\GeV$, $W_{\rm DIS-1}=1.6\GeV$, and $W_{\rm DIS-2}=1.65\GeV$.
The transition parameters given have been determined by comparison with the electroproduction data, as described in Ref.~\cite{Buss:2011mx}.
The DIS processes in the interim region above $W_{\rm DIS-1}$ and below $W_{\rm RES-2}$, where both resonance excitations and DIS are present, account for the background processes giving a few mesons in the final state  beyond the \onepion{}  background.
They also account for those (very few) resonances, which are not included in the GiBUU implementation because their electromagnetic properties are not known.
With this choice, the DIS events become noticeable at neutrino energies above about $1-2\GeV$. At this point it should be mentioned that in the absence of detailed data the cross sections in the SIS energy regime carry some uncertainties; in particular, the $2\pi$ channel, which clearly shows up in electron-induced inclusive cross sections, is completely undetermined, except for some old, low-statistics data \cite{Bottino:1969mh,Biswas:1978ey}. These uncertainties affect the total cross sections on the 3\% level.

The GiBUU method to account for nuclear effects, which is described in detail in \cite{Buss:2011mx},
essentially factorizes the
initial, primary interaction of the incoming neutrino with a target nucleon (or a pair of nucleons for 2p2h interactions) and the subsequent final-state interactions (FSI).
The target nucleons are moving inside a mean-field nuclear potential with momenta determined by the Fermi motion
in the local Thomas-Fermi approximation; all the potentials, defined in the local rest frame, have proper Lorentz-transformation properties and relativistic kinematics are used throughout. While the relativistic Fermi gas (RFG) model has difficulties to describe details of the interactions of electrons with nuclei at lower energies and energy transfers in the QE region we have shown in Refs.~\cite{Buss:2007ar,Leitner:2008ue} that at higher energies and energy transfers the GiBUU model leads to a description of QE scattering of the same quality as that reached in other more refined models of nuclear structure. We attribute this success partly to the use of a \emph{local} RFG model as well as to the use of a momentum-dependent nucleon potential that takes nuclear binding effects and final-state interactions in the first, primary interaction into account. In the energy region treated in this paper QE scattering amounts anyway only to a small part of the total cross section while pion production and DIS become more important. For these latter two processes GiBUU gives results that are very reliable when compared to data (see Figs.\ 28, 29, and 33 in Ref.~\cite{Buss:2011mx}).

Thus, we consider the cross sections of neutrino/antineutrino interactions
\[
\begin{array}{l}
\nu(k^\mu)  N (p^\mu) \to \mu^- (k'{}^\mu) X \ ,
\\[4mm]
\bar\nu(k^\mu)  N (p^\mu) \to \mu^+ (k'{}^\mu) X \ ,
\end{array}
\]
with a bound nucleon,
which has a non-vanishing three-momentum and effective mass $m_N^*$ that takes into account the nuclear potential.
In the interaction possible Pauli blocking of the final state is explicitly taken into account.
The cross sections are then obtained by summing over the interactions with all nucleons taking the flux factor $(q \cdot p)/(q_0 p_0)$, which transforms the neutrino-nucleon flux into the neutrino-nucleus flux \cite{Kulagin:1989mu}, into account.

The total absorption cross section is then calculated as
$\sigma_{\rm tot}=\sigma_{\rm QE}+\sigma_{\rm RES}+\sigma_{\rm bgr}+\sigma_{\rm DIS}$.
For QE scattering and resonance excitations the primary interaction of a neutrino with an individual target nucleon
is described taking into account the full in-medium kinematics with an energy-dependent nuclear potential
for both the incoming and the outgoing baryons. The background cross section is essentially the difference between the
experimentally known \onepion production cross section and the resonance contribution to these processes.
The details are given in Ref.~cite{Leitner:2008ue}.

High-energy electron and neutrino  data on nuclear targets, which are dominated by the DIS processes, have conventionally been analyzed by assuming that the inclusive double-differential cross sections assume the same form as those for the free reaction and all possible nuclear effects have been absorbed into the
nuclear modification of the structure functions (see, e.g.,\ Refs.~\cite{Conrad:1997ne,Kulagin:2007ju,Haider:2011qs}).
In contrast, the present model works quite differently.  The calculations explicitly contain nuclear effects such as
Fermi motion and Pauli blocking, but no intrinsic structural change of bound nucleons.
Any disagreement of our curves with the data will thus reveal ``genuine'' nuclear in-medium effects not included in the model.
For QE scattering and resonance excitations the effects of the nuclear
potential are explicitly taken into account by using spectral functions with the properly shifted real parts.
GiBUU also allows us to account for the in-medium width modification of the resonances;
in the present high-energy calculations, however, we neglect this effect for simplicity and because its influence on the observables calculated here is small.

\textsc{pythia}, used to generate the DIS events, is a free-nucleon generator, which does not allow explicit use of a nuclear potential.
We try to account for this and thus determine the elementary cross section appropriate for a bound nucleon by
adjusting the  input kinematical variables for \textsc{pythia}.
Various prescriptions to do this have been discussed in Ref.~\cite{Buss:2011mx}. We will come back to this point later in Sec.~\ref{nuclearDIS}. Here we just mention that the analogous uncertainty also exists in the standard treatment of electron- and neutrino-induced inclusive cross sections where one has to decide at which kinematical variables $\omega$ (energy transfer) and $Q^2$ (four-momentum transfer) the structure functions have to be read off
\cite{Bodek:1980ar,DeForest:1983vc,Koltun:1994zd}.

After being produced in the initial interaction, outgoing hadrons propagate throughout the nucleus.
In GiBUU this process of FSI is modeled by solving the semi-classical
Boltzmann-Uehling-Uhlenbeck equation, which, again, respects relativity \cite{Buss:2011mx}.
It describes the dynamical evolution of the one-particle phase-space density for each particle species
under the influence of the mean field potential, introduced in the description of
the initial nucleus state. Equations for various particle species are coupled through this mean field and
also through the collision term. This term explicitly accounts for changes in
the phase-space density caused by elastic and inelastic collisions between particles.
We note here that, contrary to some other uses of this term, we call FSI all the interactions that take place after the initial, primary reaction while the hadrons propagate through the nucleus. However, the same potential that governs this later evolution is also present during this first reaction and affects its outcome, mainly due to the energy dependence of the mean-field potential that changes the final-state phase space.

At higher energies and, in particular, higher $Q^2$, formation times and color transparency phenomena may become important. 
Relevant data were taken by EMC and HERMES collaborations some years ago.
In Ref.~\cite{Gallmeister:2007an} we have analyzed these data in the energy regime from about 10 to 200 GeV and have found that only a linear increase of the prehadronic cross sections with time fits both sets of data simultaneously. Therefore, the quantum-diffusion model of Farrar et al.\ \cite{Farrar:1988me} is implemented in GiBUU. At the lower energies formation times are determined by the lifetimes of resonances, which are explicitly propagated, and as such are automatically included in GiBUU.

\section{Results on the nucleon  \label{nucleon}}

\begin{figure}[htb]
\includegraphics[width=\columnwidth]{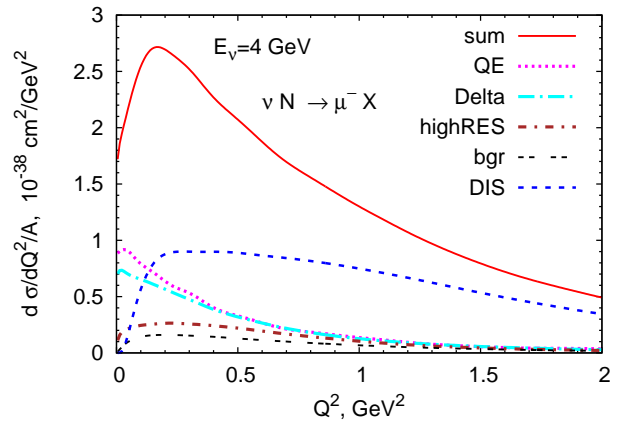}
\caption{(Color online) Cross section $\dd \sigma/\dd Q^2$ per nucleon for CC neutrino scattering off an isoscalar target
 for $E_{\nu}=4\GeV$.}
\label{fig:isoscalar-dsidQ2}
\end{figure}

At low neutrino energies up to about 2 GeV, the cross sections for neutrino reactions on the nucleon
were presented in Ref.~\cite{Leitner:2006ww} and showed good agreement with the experimental data.
At higher energies (above  about $E_\nu \approx 2-3 \GeV$) QE, resonance ($\Delta$ and higher resonances) and background (bgr)
contributions practically do not change with energy. This is because they strongly fall off with increasing $Q^2$, as shown
in Fig.~\ref{fig:isoscalar-dsidQ2} for the example of an isoscalar target for a neutrino energy of $4\GeV$.
The DIS contribution, on the other hand, is much flatter in $Q^2$; its absolute value will also grow with increasing energy.

\begin{figure*}[htb]
\begin{minipage}[c]{0.48\textwidth}
\includegraphics[width=\textwidth]{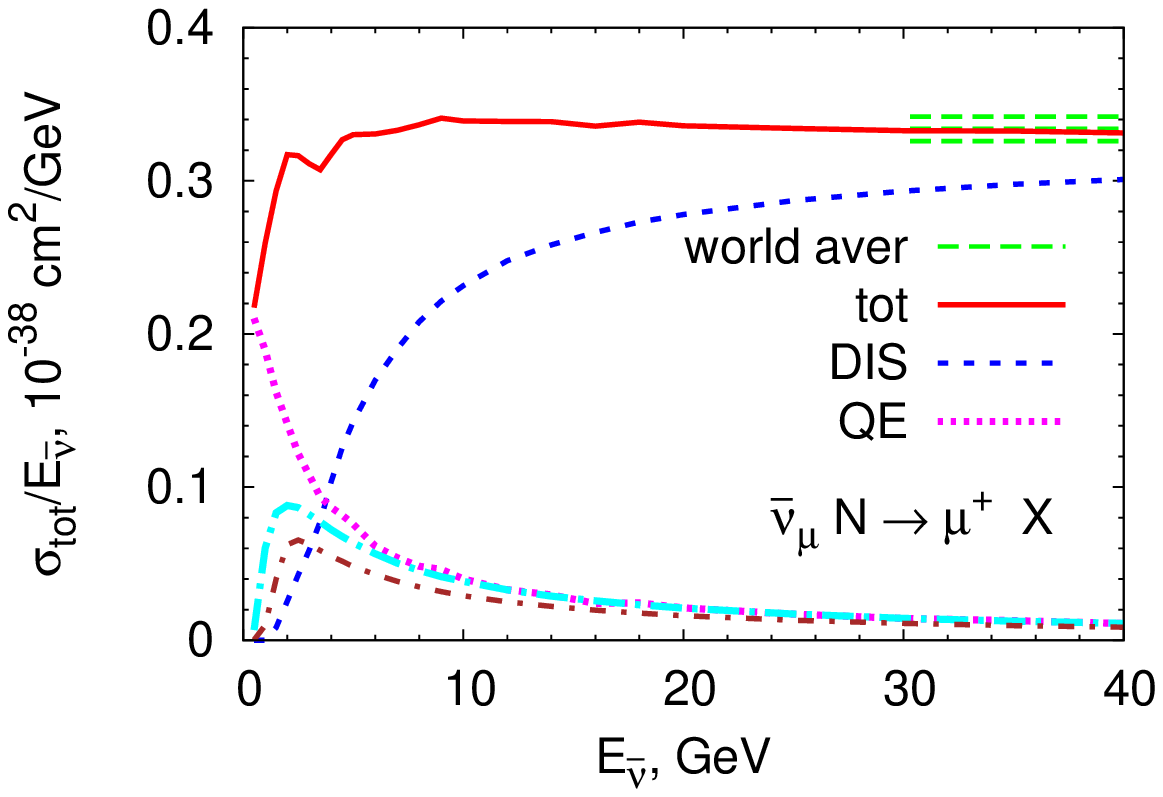}
\end{minipage}
\hfill
\begin{minipage}[c]{0.48\textwidth}
\includegraphics[width=\textwidth]{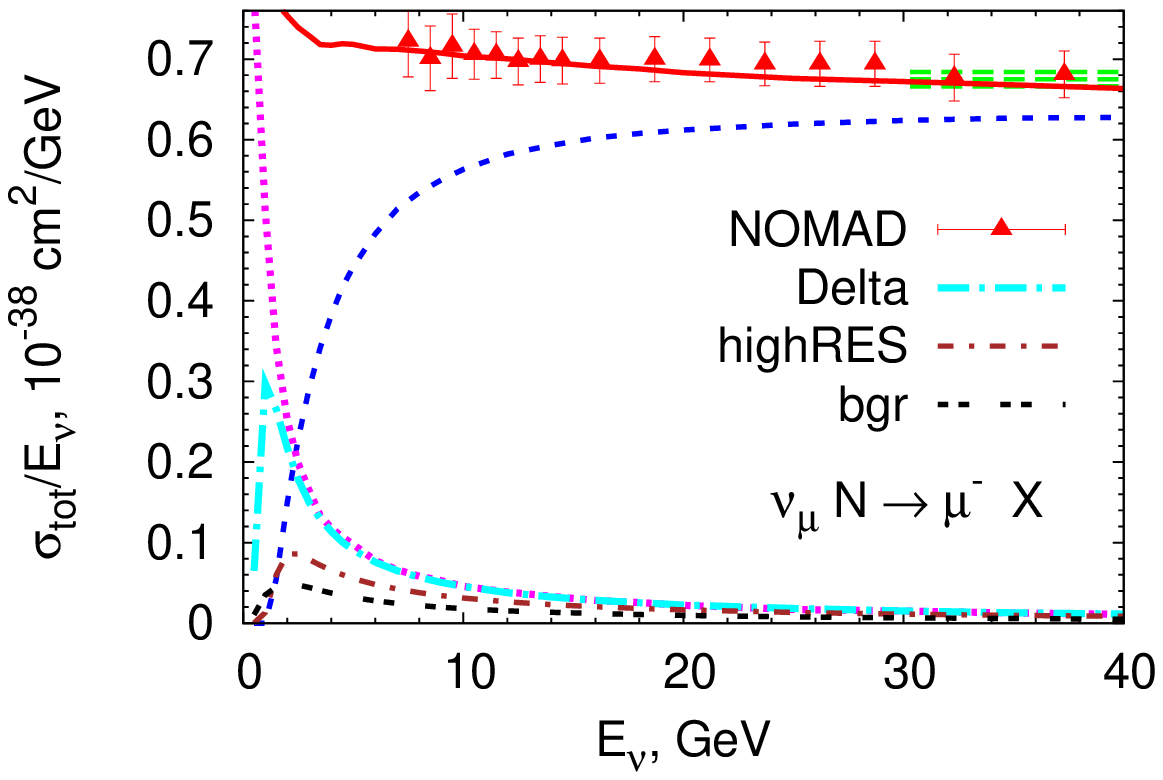}
\end{minipage}
\caption{(Color online) Total (red solid), DIS (blue dashed) and other cross sections per energy per nucleon  for antineutrino (left) and neutrino (right)
inclusive scattering on an isoscalar target. Curves are compared with the world averaged values, denoted by the dashed band.
Data points from the NOMAD experiment \cite{:2007rv} are shown as solid triangles.}
\label{fig:compare-isoscalar-free}
\end{figure*}

Thus, at neutrino energies above a few GeV the major contribution to the total cross section comes from DIS. According to the predictions of the parton model, the DIS cross section grows linearly with energy. At high neutrino energies the data are, therefore, conveniently presented as cross section
per energy $\sigma_\textrm{tot}/E_\nu$.

This cross section was measured starting in the 1980s by the CCFR, BEBC, CHARM, IHEP, NuTeV, CHORUS and NOMAD collaborations; see Ref.~\cite{Amsler:2008zzb} for the plot with all data and references. The world average values derived for $E_\nu > 30 \GeV$ are given for an isoscalar target:
$\sigma_\textrm{tot}^{\nu}/E_\nu = 0.667\pm 0.014 \cdot 10^{-38} \cm^2/\mathrm{GeV}$ for neutrinos
and $\sigma_\textrm{tot}^{\bar\nu}/E_{\bar\nu} = 0.334\pm 0.008 \cdot 10^{-38} \cm^2/\mathrm{GeV}$ for antineutrinos.

Figure~\ref{fig:compare-isoscalar-free} shows the results of our calculations of the total cross section
as well as the DIS and other contributions indicated in the figure for both neutrino and antineutrino scattering
on an isoscalar target. The world average values and their error bands are indicated for comparison for $E_\nu > 30 \GeV$.
The NOMAD experiment has recently performed measurements on composite target with a measured composition of $52.43\%$ protons
and $47.57\%$ neutrons \cite{:2007rv}. The measurements were then corrected for the nonisoscalarity, and the results are
presented as isoscalar cross section. The data points are shown in Fig.~\ref{fig:compare-isoscalar-free} as solid triangles.

For neutrinos, the DIS contribution becomes larger than the $\Delta$ contribution
at about 3 GeV, and at 5 GeV it is already about $60\%$ of the total and reaches $95\%$ at higher energies.
The rest is to be attributed to other channels.
For antineutrinos,
the DIS contribution becomes larger than the $\Delta$ channel at about 4 GeV; it is $40\%$ of the total at $E_{\bar\nu}\sim 5\GeV$ and reaches $95\%$ at higher neutrino energies. This clearly has
implications for theoretical analyses of the cross section measurements in the MINOS and NO$\nu$A experiments.

The dip in the total antineutrino cross section at $E_\nu\sim 3-4 \GeV$ (there is an indication for a similar effect in the neutrino cross section) shows up as such only in the $1/E_\nu$ scaled cross section. It is caused by an interplay of the downfall of the resonance contributions and the rise of DIS. It could indicate that our model misses some strength here. This intermediate energy region would be most sensitive to the pion (one or more) background. A clarification must wait until much more precise data for $1\pi$ and $2\pi$ production become available.

The figure also shows for both neutrinos and antineutrinos a slight decrease of the calculated $\sigma_{\rm tot}/E$ at large energies. In our calculations this slow decrease is due to the quasielastic and resonance contributions that fall with energy. The high-energy data also exhibit a small downward slope usually attributed to  perturbative QCD and heavy quark effects. This latter slope was not taken into account in deriving the world average values, where it was assumed to be negligible \cite{Conrad:1997ne}.

\section{Quasielastic Scattering}

\begin{figure}[bht]
\centering
\includegraphics[width=0.9\columnwidth]{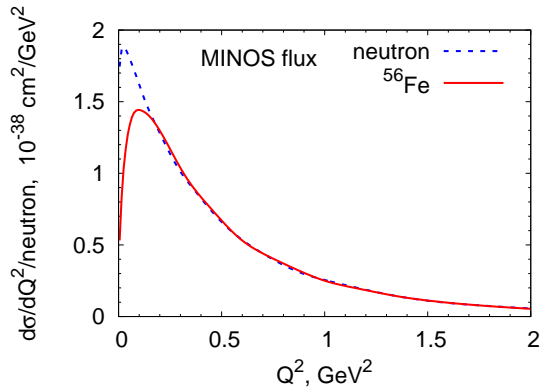}
\caption{(Color online) Cross section $\dd \sigma/\dd Q^2$ per neutron for neutrino CC QE scattering as a function of
squared four-momentum transfer. The cross section is averaged over the MINOS flux.}
\label{fig:MINOSflux-QE-dsidQ2}
\end{figure}

As we have shown in Fig.~\ref{fig:compare-isoscalar-free}, with increasing neutrino energy the CC QE cross section gives smaller and smaller relative contribution to the total cross section. However, this channel is still of importance because of its special role in the reconstruction of the neutrino energy. The MINOS \cite{Mayer:2011zz}, Miner$\nu$a \cite{McFarland:2011jd}, and ArgoNeuT \cite{Spitz:2010dj} experiments all aim at studying this channel. They use the same neutrino flux, but have different detectors and different criteria for selection of QE events, which, again, differ from those of low-energy experiments like MiniBooNE and K2K. 

Here we present in Fig.~\ref{fig:MINOSflux-QE-dsidQ2} only the QE cross section as a  function of $Q^2$, calculated for an axial mass of $M_A = 1 GeV$. The plot is shown for the MINOS flux; the results obtained with the NO$\nu$A flux are nearly indistinguishable, as expected for neutrino energies above $1.5\GeV$. The elementary QE cross section used here is given in Ref.~\cite{Leitner:2006ww}. The dashed line shows the free cross section on a single neutron; the solid line takes into account nuclear effects in iron. Those reveal themselves only at low momentum transfer; for $Q^2>0.2 \GeV^2$ the ``neutron'' and ``${}^{56}$Fe'' curves are practically indistinguishable.

We show the flux-integrated QE result here in order to illustrate that the $Q^2$ distribution for QE scattering looks very similar to that at the lower-energy experiment MiniBooNE. Comparison with Fig.\ \ref{fig:isoscalar-dsidQ2} shows that the different reaction mechanisms have quite different $Q^2$ behavior. Whereas both QE and the $\Delta$ contribution fall off quite quickly with $Q^2$, driven by the $Q^2$ dependence of their form factors, the DIS contribution is, after an initial rise, rather flat out to larger momentum transfers.

We stress here that the solid curve in Fig.\ \ref{fig:MINOSflux-QE-dsidQ2} depicts the true CC QE cross section without any initial pion events and without any initial 2p-2h interactions. The latter would contribute not to QE scattering, but to the inclusive pionless cross section. Taking the MiniBooNE results for guidance they would contribute about 30\% to the latter, but their overall influence on the total cross sections would be negligible at the higher energies where the dominating DIS cross section goes $\propto E_\nu$.

\section{Nuclear effects in DIS  \label{nuclearDIS}}

\begin{figure*}[b!ht]
\begin{minipage}[c]{0.48\textwidth}
\includegraphics[width=\textwidth]{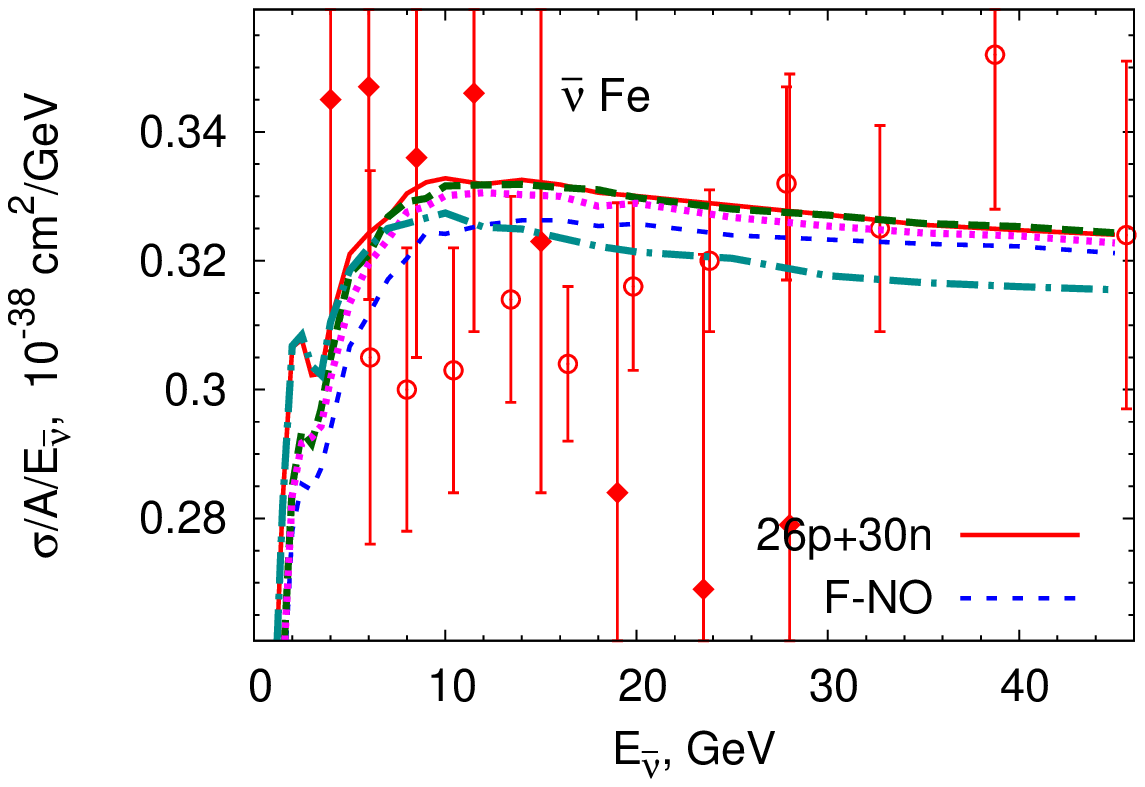}
\end{minipage}
\hfill
\begin{minipage}[c]{0.48\textwidth}
\includegraphics[width=\textwidth]{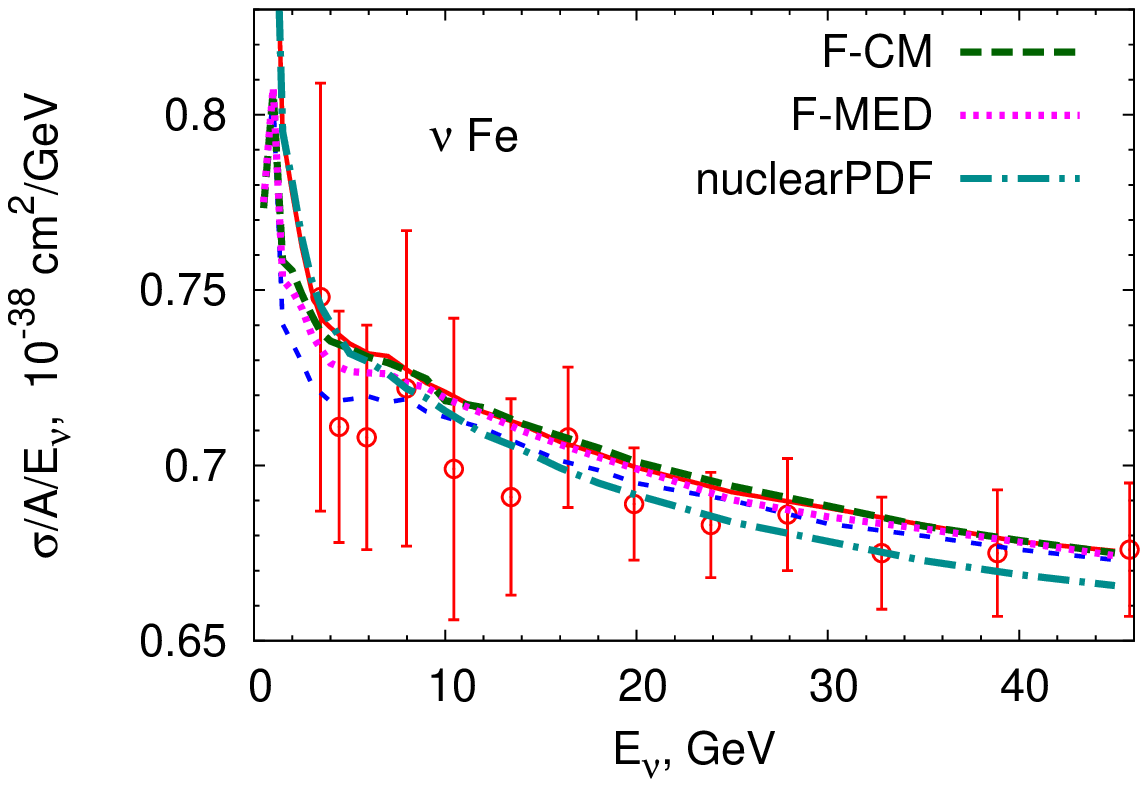}
\end{minipage}
\caption{(Color online) Cross sections $\sigma_{\rm tot}/E_\nu$ per nucleon for antineutrino  $\bar\nu_\mu \text{Fe} \to \mu^+ X$ (left)
and neutrino $\nu_\mu \text{Fe} \to \mu^- X$ (right) inclusive scattering off iron.
Experimental data are taken from Refs.~\protect\cite{Adamson:2009ju} (MINOS, open circles) and 
 \protect\cite{Anikeev:1995dj} (IHEP-JINR, solid diamonds).
}
\label{fig:compare-nuclear}
\end{figure*}

All the neutrino experiments, used to derive the world average cross sections on the nucleon
mentioned in the previous section, were actually performed
on nuclear targets. If the target was nonisoscalar, a corresponding correction
was introduced and the  isoscalar cross section was extracted \cite{Conrad:1997ne}.
Such a procedure as well as the introduction of the ``world-average'' value is meaningful
only if nuclear corrections are very small.

The EMC effect shows that there are nuclear corrections to the free cross sections for electrons \cite{Piller:1999wx}.
For neutrino reactions, however, the situation is controversial.
On one hand, nuclear parton distributions, based on electromagnetic scattering data and intended for description
of both charged lepton and neutrino reactions, were introduced. For a review and a list
of recent parametrization see, for example, Ref.~\cite{Hirai:2009mq}.
On the other hand, a recent investigation \cite{Schienbein:2007fs,Kovarik:2010uv} showed that in neutrino reactions
nuclear corrections to parton distributions have about the same magnitude as for electrons, but
have a very different dependence on the Bjorken-$x$ variable.
The topic remains controversial, with the hope that future precise Miner$\nu$a results on various
targets will clarify the situation.

As we have already mentioned above, the GiBUU code uses \textsc{pythia} for the simulation of the DIS processes.
Since the \textsc{pythia} code was designed for elementary reactions
we have to provide some ``quasi-free'' kinematics as input to \textsc{pythia} that removes the effects of the binding potential on the nucleon.
Various prescriptions to do this  have been used (for details see Ref.~\cite{Buss:2011mx}) and are
compared with each other and with the free cross section in Fig.~\ref{fig:compare-nuclear}.
The corresponding cross sections are denoted as ``F-NO'' (the invariant energy $W^2$ of the boson-nucleon system
is calculated as $(k-k'+p)^2$ and not corrected),
``F-CM'' (bound nucleon is boosted into the center-of-momentum frame and the nuclear potential is removed
from its energy, the nucleon then is boosted back, and $W$ is calculated), and
``F-MED'' ($W^2$ is taken as  $(k-k'+p)^2 - m_N^*{}^2 + m_N^2$).
In all these calculations parton distributions appropriate for free nucleons have been used.

It is seen that the difference between these various prescriptions is quite small (about 2 \% at the lowest energy and less at the highest energy); also, the results all approach the free cross section at the highest energy. We consider small differences between the results obtained with the various prescriptions mentioned above as intrinsic uncertainty of the GiBUU code, reflecting the lack of a detailed understanding of nuclear effects. No other event generator, as far as we know, accounts for nuclear corrections in high--energy neutrino reactions. Nuclear parton distribution functions
from Ref.~\cite{Eskola:1998df} are also implemented as one of the options to use.
To avoid double counting, nuclear potential and Fermi motion are switched off in such calculations.
The result (``nuclearPDF'') as well as the free cross section for iron composition
(``26p+30n'') are also shown in Fig.~\ref{fig:compare-nuclear}.

\begin{figure}[b!ht]
\includegraphics[width=\columnwidth]{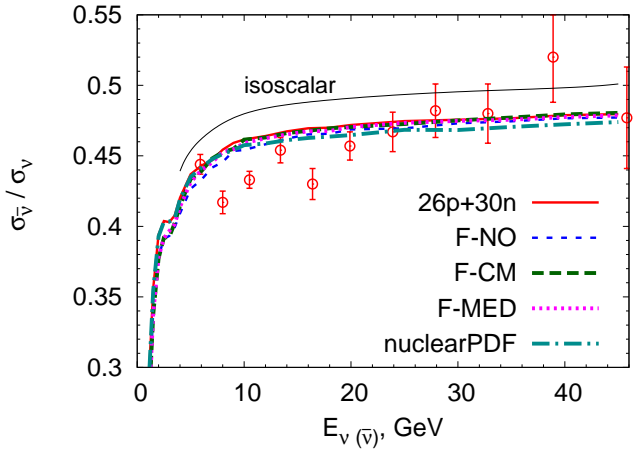}
\caption{(Color online) Ratio of the antineutrino to neutino cross sections in scattering off iron. Also shown is the calculated ratio for an isoscalar-corrected target.
Experimental data are taken from Ref.~\protect\cite{Adamson:2009ju} (MINOS, open circles).}
\label{fig:compare-nuclear-nu-barnu-ratio}
\end{figure}

Figure~\ref{fig:compare-nuclear} shows, that for antineutrinos our curves are  within the spread of the MINOS and IHEP-JINR data \cite{Anikeev:1995dj}.
The overall agreement of our calculations with the data is,  therefore, better than the agreement of the data with each other.
At low energies, where the main contribution comes from the QE and resonance production, nuclear effects are known to reduce the
neutrino and antineutrino cross section.
This is why,  at $E_{\bar\nu}<5\GeV$, the curves ``F-NO'', ``F-CM'', and ``F-MED'' that take into account the nuclear effects explicitly
lie noticeably lower than the ``26p+30n'' curve. At high energies, however, the curves converge toward each other.
The ``nuclearPDF'' curve, which implies modification of the DIS channel only, coincides with the ``26p+30n'' curve at low energies
and consistently deviates from it to lower values of the cross section at higher energies where DIS dominates.
The peak and dip in the region $3-4\GeV$ in the free cross section have the same origin as for the isoscalar cross section,
as discussed in the previous section. Nuclear effects, mainly Fermi motion, wash out this structure so it is no longer visible in the nuclear
cross sections.

For neutrinos our curves are in good agreement with the recent MINOS experiment, they lie within the errors of the data points.
For both neutrinos and antineutrinos nuclear effects are noticeable at low energies. For higher energies, $E_{\nu}>5\GeV$, the curves ``F-NO''
``F-CM'' and ``F-MED'' all approach each other and the ``26p+30n'' curve, reflecting the expected disappearance of nuclear effects with increasing energy.

In Fig.\ \ref{fig:compare-nuclear-nu-barnu-ratio} we show the ratio of antineutrino cross section to the neutrino one as a function of energy, which is in good
agreement with the recent MINOS data.
The ratio rises with energy, gradually flattening out.  It is  below its asymptotic value obtained for an isoscalar-corrected target ($\approx 0.5$ at high energies \cite{Conrad:1997ne}); the corresponding isoscalar  curve is also shown in the figure. It is interesting that this ratio is remarkably insensitive to any nuclear effects, even at the lower energies, as illustrated by the fact that now all the curves for the various in-medium correction methods lie essentially on top of each other.


\section{Semi-Inclusive channels \label{semi-inclusive}}

In this section we present spectra for outgoing pions and nucleons.
All calculations have been done for the MINOS flux on an iron target and the NO$\nu$A flux on a carbon target.
As emphasized, for example, in Ref.~\cite{Leitner:2010kp},
acceptance cuts or detector thresholds can have a significant and non-trivial influence on the measured values.
In present calculations, no such cuts and no detector thresholds are assumed for the outgoing hadrons.

In many cases, FSI significantly modify
the shapes of the final particle spectra. Particles get slowed down in the medium by collisions with other nucleons leading to a pileup of cross section at low kinetic energies.
Such modification is seen, for example,
in photopion production~\cite{Krusche:2004uw} and is well described by GiBUU.
A similar change should be observed in neutrino reactions. In the following we now discuss nucleon knock-out, pion production, and strangeness production in neutrino-induced reactions.

\subsection{Nucleon kinetic energy distributions}

\begin{figure*}[hbt]
\hfill
\includegraphics[width=0.6\textwidth]{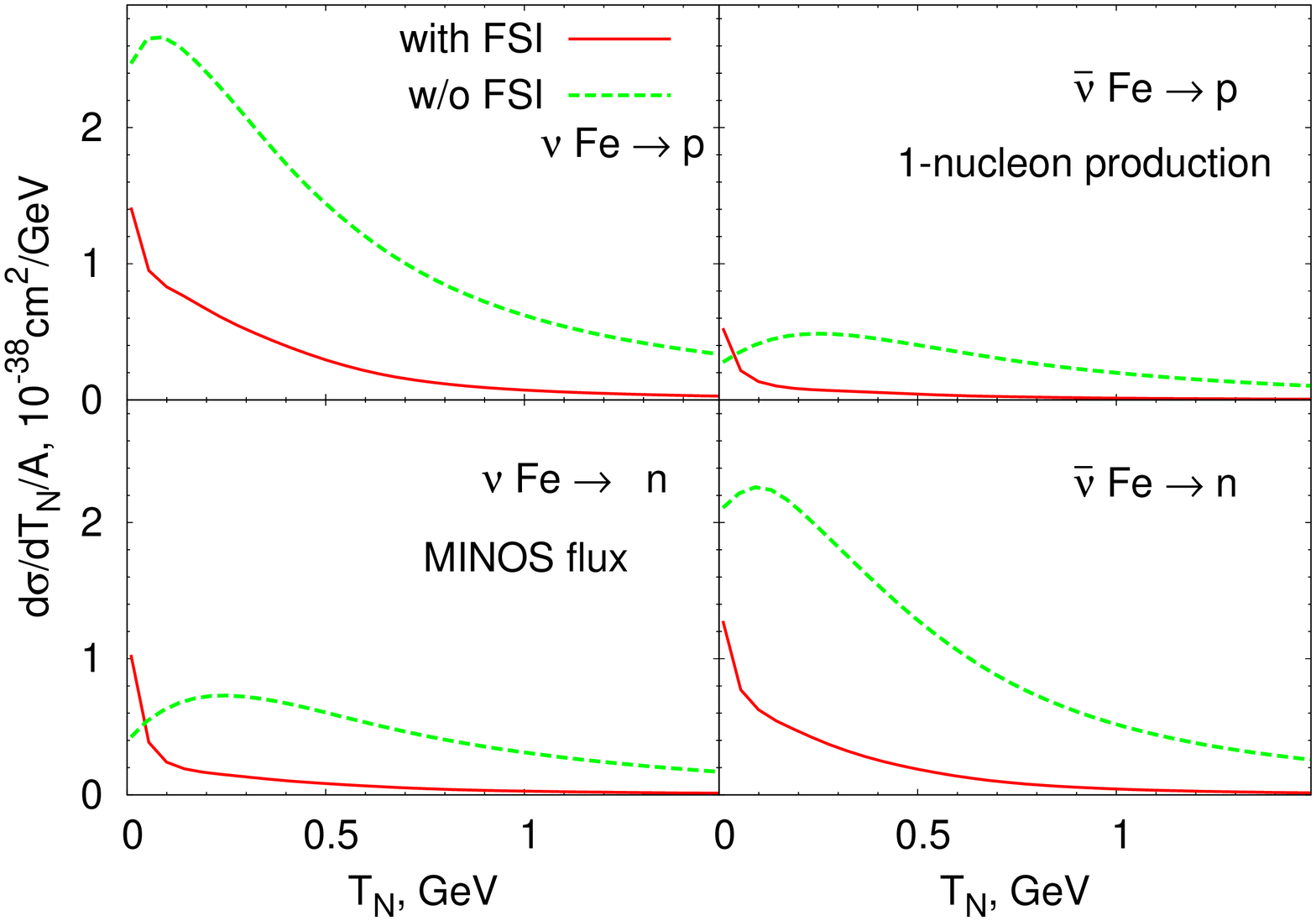}
\hfill
\includegraphics[width=0.33\textwidth]{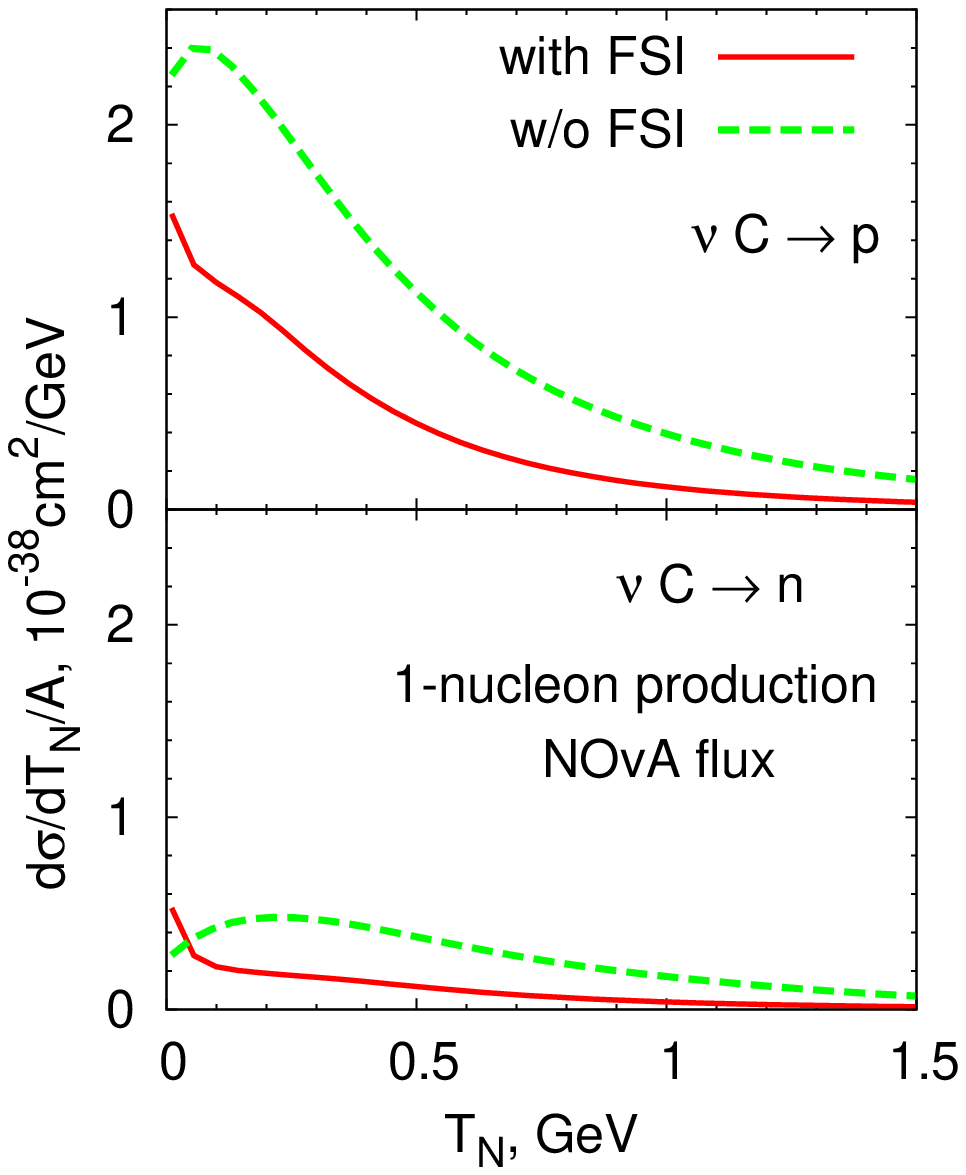}
\hfill
\caption{(Color online) Kinetic energy distributions  per target nucleon  for \onenucleon
(one nucleon of the indicated charge and no other nucleons) production
in neutrino and antineutrino scattering off iron and carbon.
The left block of figures shows the results obtained for the MINOS flux, and the right block those for the NO$\nu$A flux. }
\label{fig:MINOS-ekin-with-wo-FSI-1-nucleon}
\end{figure*}

\begin{figure*}[hbt]
\hfill
\includegraphics[width=0.60\textwidth]{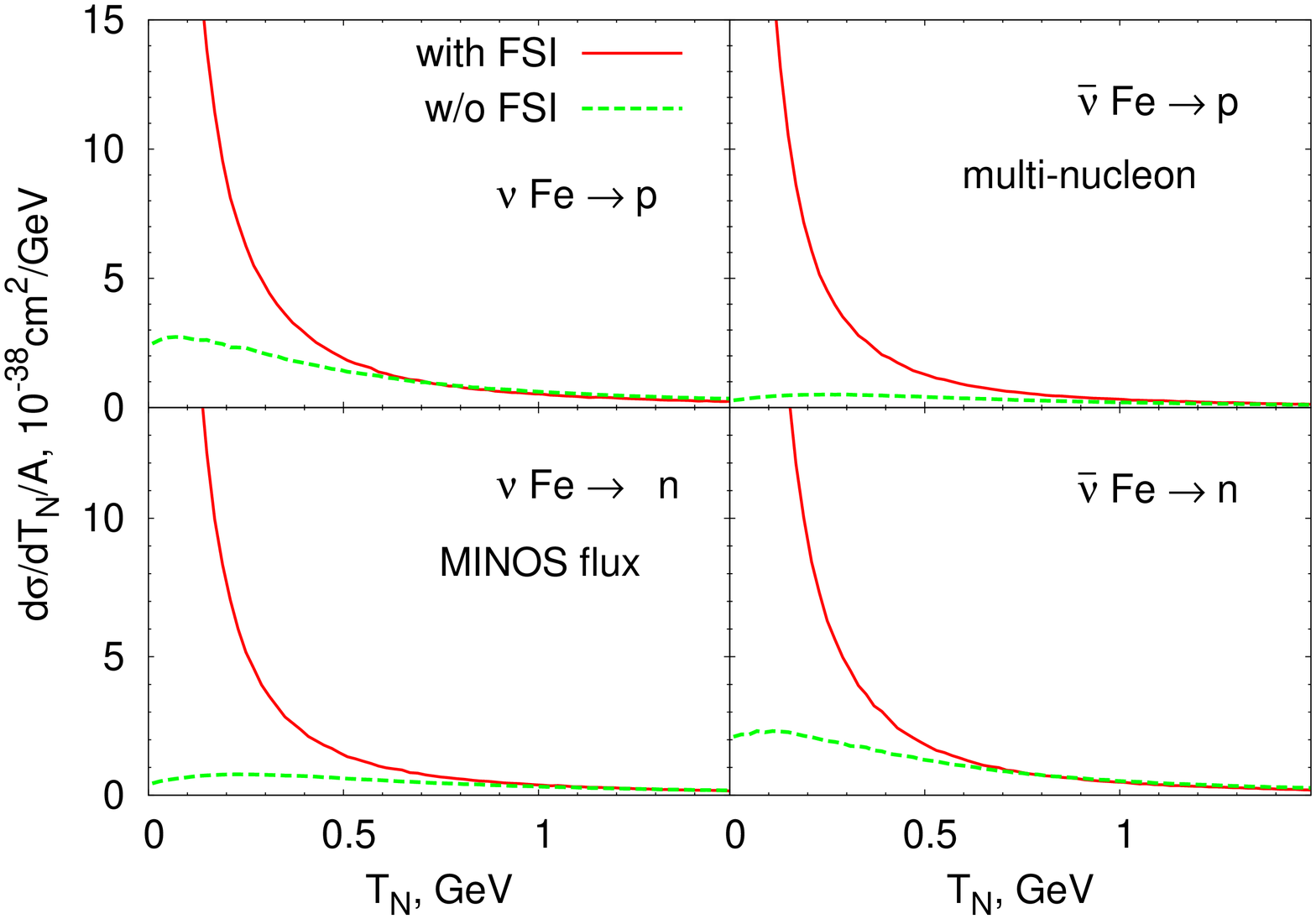}
\hfill
\includegraphics[width=0.33\textwidth]{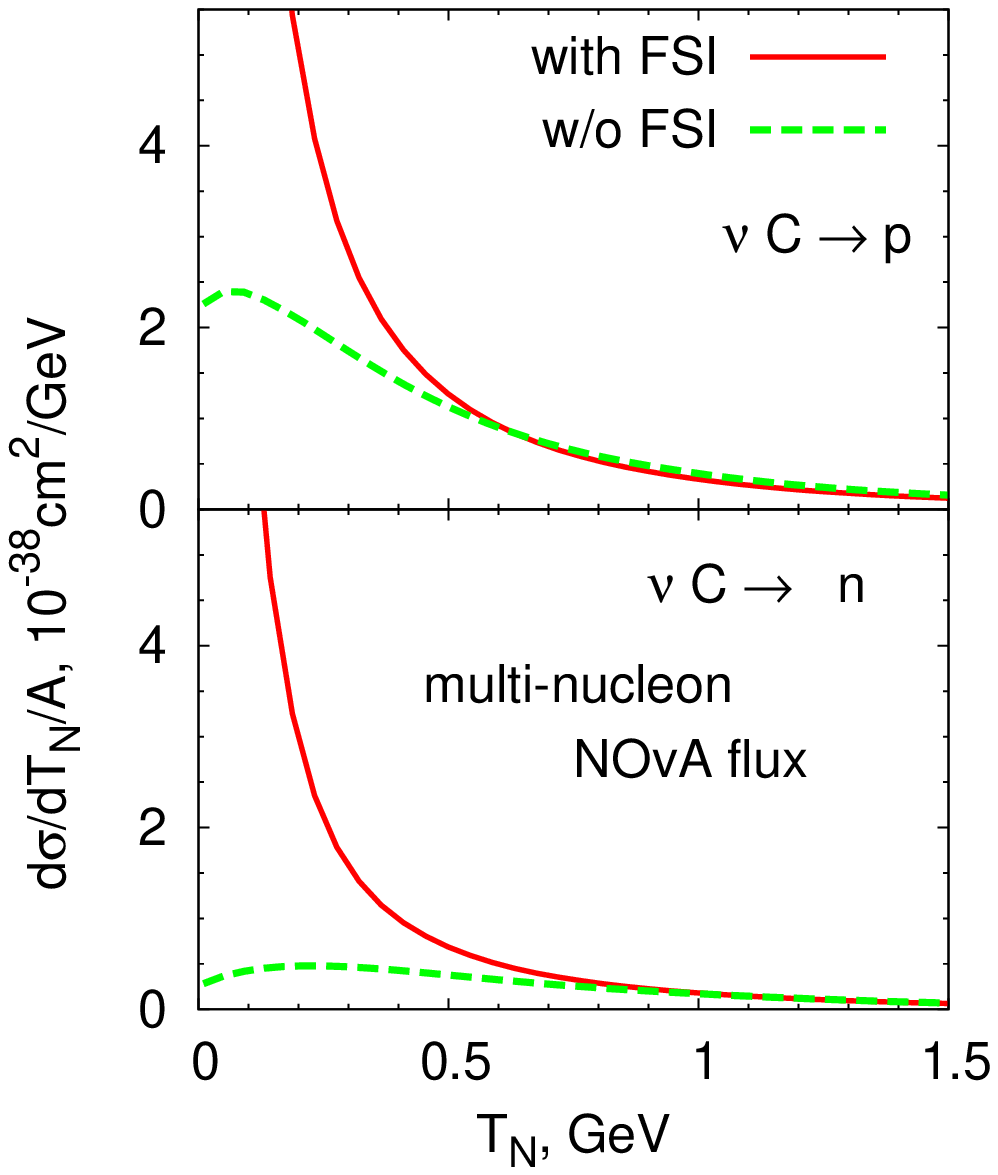}
\hfill
\caption{(Color online) Kinetic energy distributions  per target nucleon for multi-nucleon
(at least one nucleon of the indicated charge and any number of other nucleons) production
in neutrino and antineutrino scattering off iron and carbon. }
\label{fig:MINOS-ekin-with-wo-FSI-nucleon-MULTI}
\end{figure*}

Knock-out nucleons may give an insight into the actual primary interaction process and, in particular, into
the question if the impulse approximation is sufficient or if multi-nucleon interactions are essential.
The kinetic energy distributions of outgoing nucleons are shown in Fig.~\ref{fig:MINOS-ekin-with-wo-FSI-1-nucleon}
for \onenucleon events (one nucleon of a given charge and no other nucleons in the final state)
and in Fig.~\ref{fig:MINOS-ekin-with-wo-FSI-nucleon-MULTI} for multi-nucleon events
(at least one nucleon of the indicated charge and any number of other nucleons).

The decrease at higher energies ($T > 0.05\GeV$) due to FSI is natural to expect, because a nucleon can
rescatter in the nucleus and knock out another nucleon; the nucleon is then gone from the \onenucleon channel. At the same time, its kinetic energy would be spread between the two secondary nucleons. If these two have an energy large enough, they could, in turn, produce more lower-energy nucleons. The same knock-out can be caused by a pion produced in a primary interaction. This process, which can develop as a cascade, leads to an increase of multi-nucleon events with low kinetic energies.

This is indeed seen in Fig.~\ref{fig:MINOS-ekin-with-wo-FSI-nucleon-MULTI} which shows the kinetic energy spectra for multi-nucleon knock-out for both protons
and neutrons. Here the cross sections before FSI are quite similar to those in the \onenucleon  events, but there is a dramatic effect of FSI with a huge increase of the cross section at smaller energies ($T_N \lesssim 0.05$ GeV).
This steep rise at smaller energies was already observed earlier in Refs.~\cite{Nieves:2005rq,Leitner:2006ww} and is due to the energy loss connected with multiple scattering as just discussed. We note that this behavior can only be obtained in a transport-theoretical treatment of the neutrino-nucleus interaction; models that work with an optical potential for the outgoing nucleons can usually only describe their loss of flux, but not where this flux goes. An exception is a treatment based on Feshbach's projection formalism, that, in principle, is equivalent to a full coupled-channels treatment and, thus, allows also a description of inelastic processes and energy loss. Applications of this formalism have so far been restricted to inclusive cross sections \cite{Meucci:2012yq} and to exclusive proton knock-out reactions \cite{Meucci:2001qc}.

For NO$\nu$A flux calculations, which are done for carbon nucleus, the picture is similar. Since the nucleus is smaller, the hadrons can leave it earlier
and the cascade processes have less time to develop.  So, FSI lead to lower nucleon multiplicities (as compared to iron) and correspondingly higher kinetic energies.  As a consequence, for \onenucleon events, the suppression is smaller than that for iron and  for multi-nucleon events the rise at low energies
is not so steep.

At this point it is in order to comment briefly on the possible
effects of initial many-body (so-called 2p-2h) excitations that are not contained in the calculations discussed here. The presence
of such excitations, which is well established in inclusive electron scattering in the region between the QE peak and the first
resonance ($\Delta$) excitation, has recently been invoked to explain the difference between the QE-like cross sections measured in
the MiniBooNE experiment and theoretical calculations \cite{Martini:2011wp,Nieves:2011yp,Barbaro:2011st,Lalakulich:2012ac}. Here we note that for the MINOS flux
most of the multi-proton knock-outs come from DIS events and that already QE plays only a minor role. Any 2p-2h contributions thus necessarily  would be quite small and can be neglected.

\subsection{Pion kinetic energy distributions}

\begin{figure*}[bht]
\centering
\hfill
\includegraphics[width=0.6\textwidth]{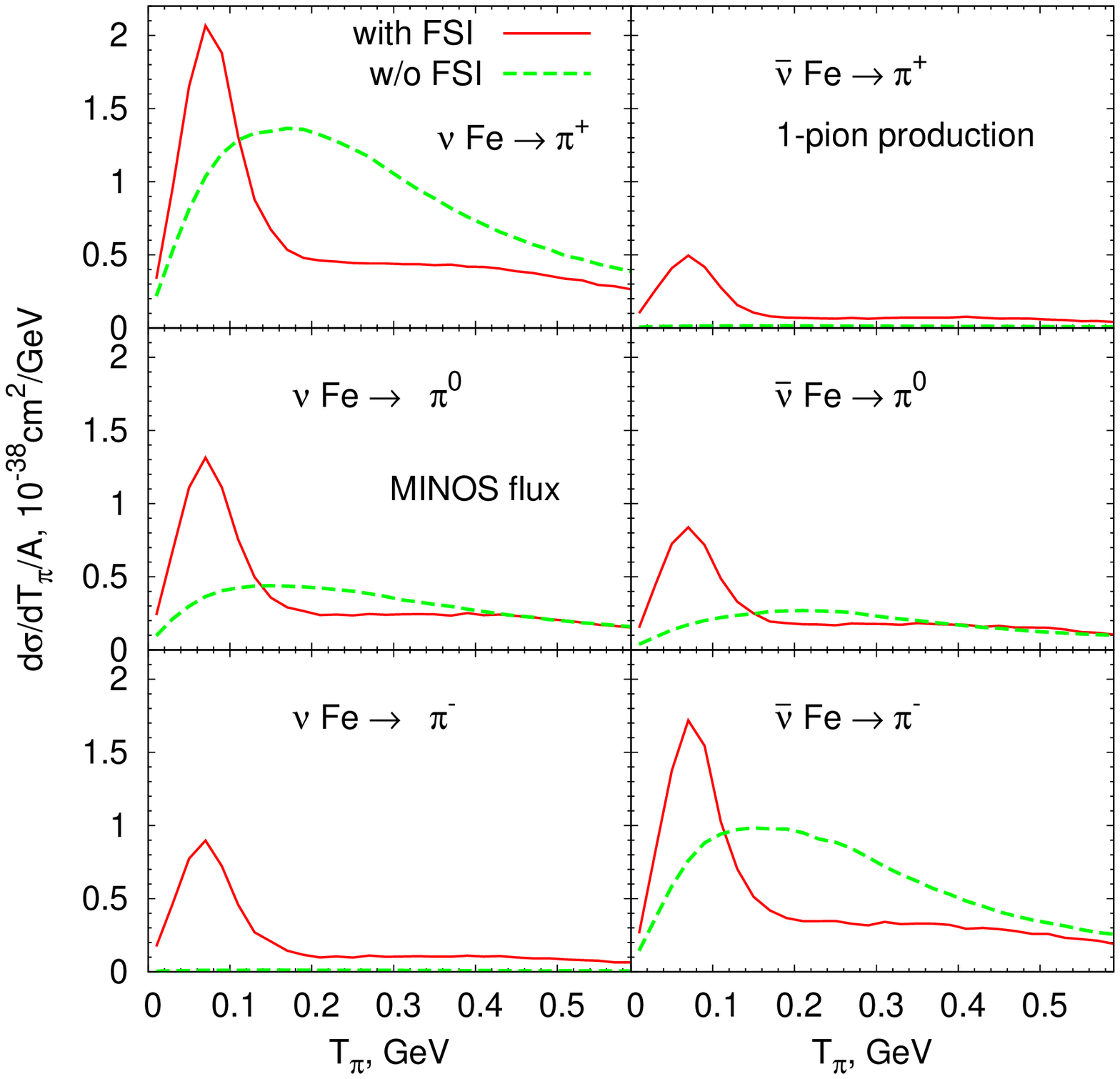}
\hfill
\includegraphics[width=0.33\textwidth]{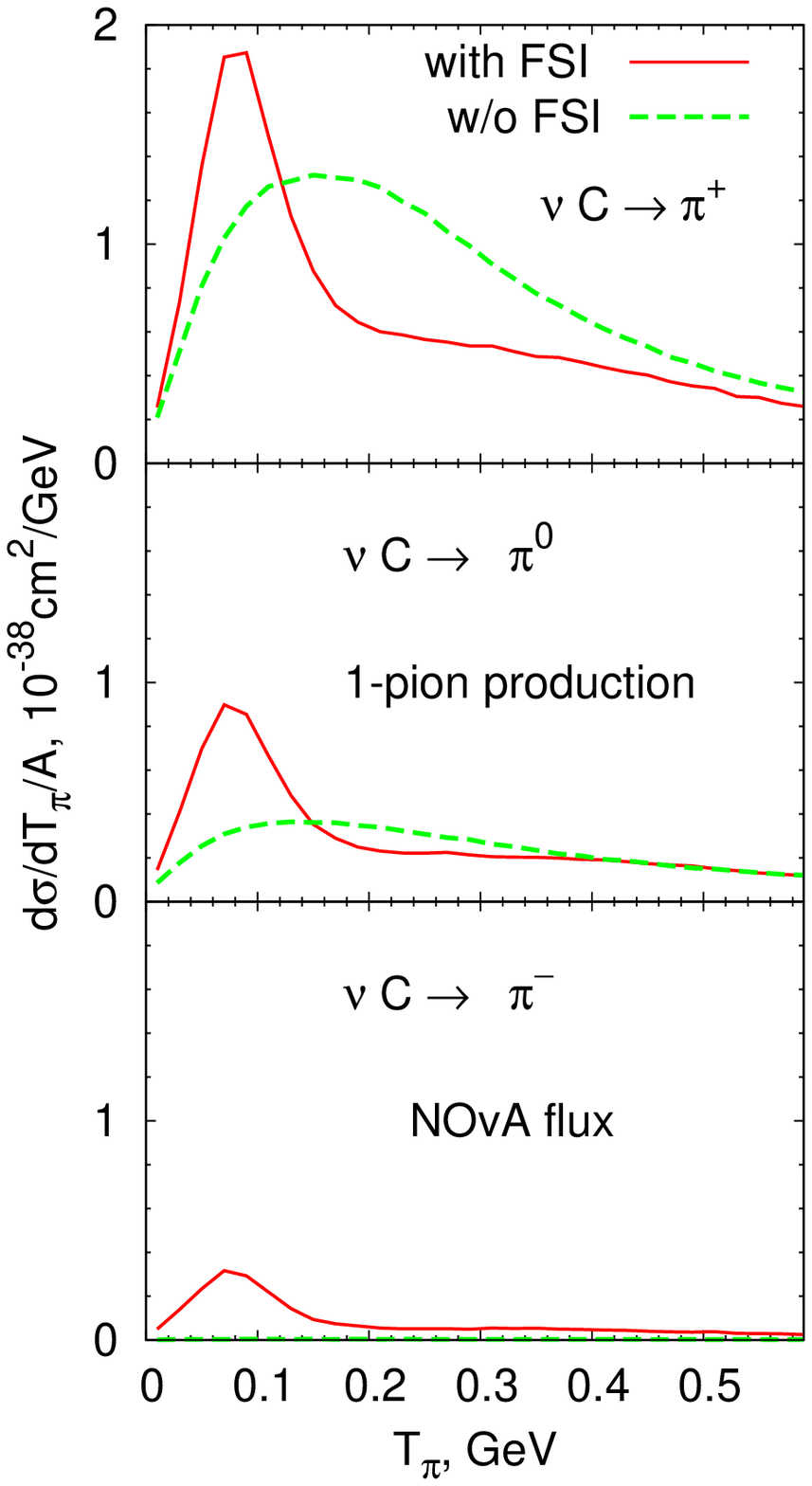}
\hfill
\caption{(Color online) Pion kinetic energy distributions per target nucleon
for neutrino and antineutrino scattering off iron and carbon for \onepion production
(one pion of the indicated charge and no other pions are produced).}
\label{fig:MINOS-ekin-with-wo-FSI-1-pion}
\end{figure*}

\begin{figure*}[t]
\centering
\hfill
\includegraphics[width=0.6\textwidth]{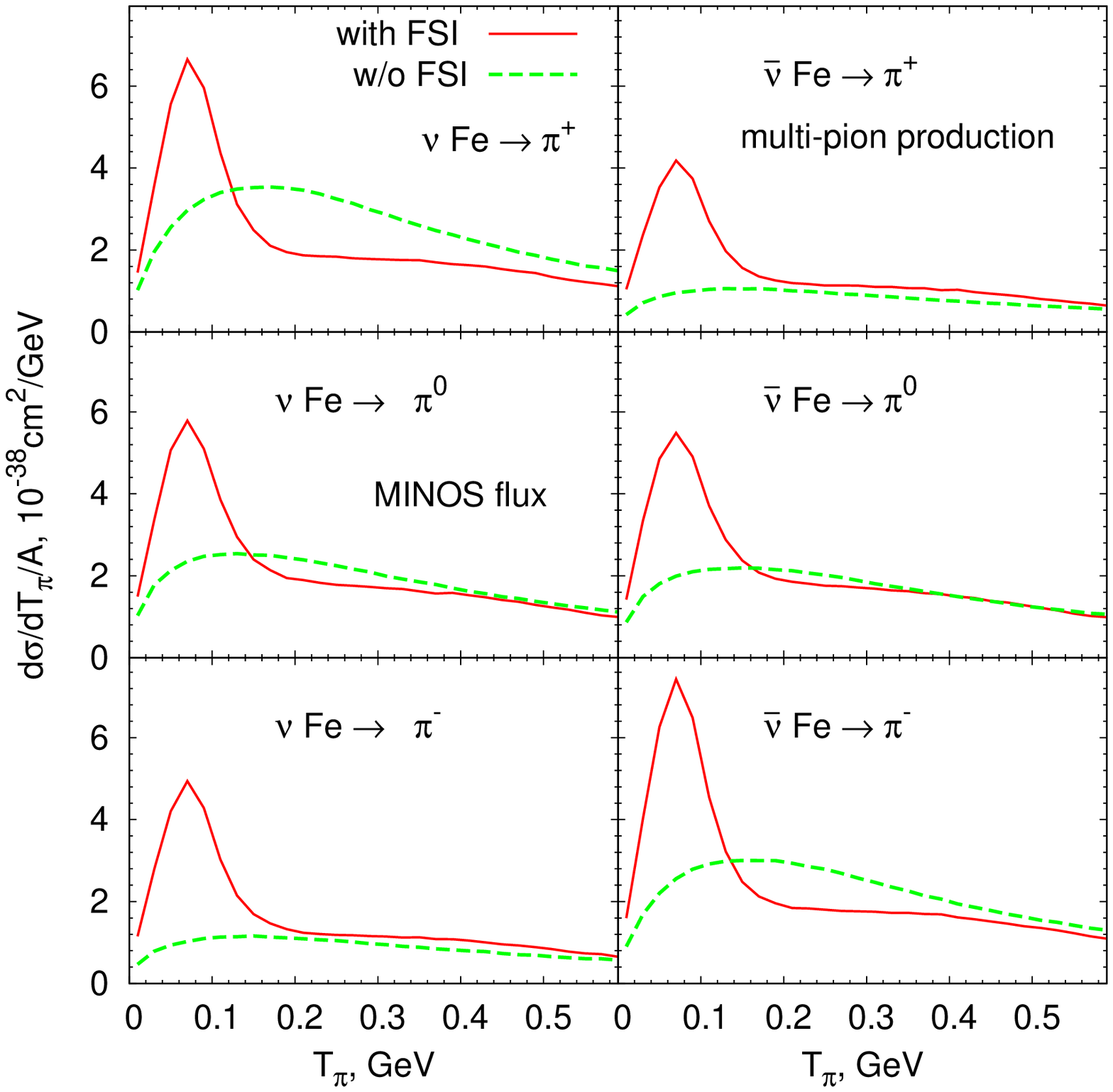}
\hfill
\includegraphics[width=0.34\textwidth]{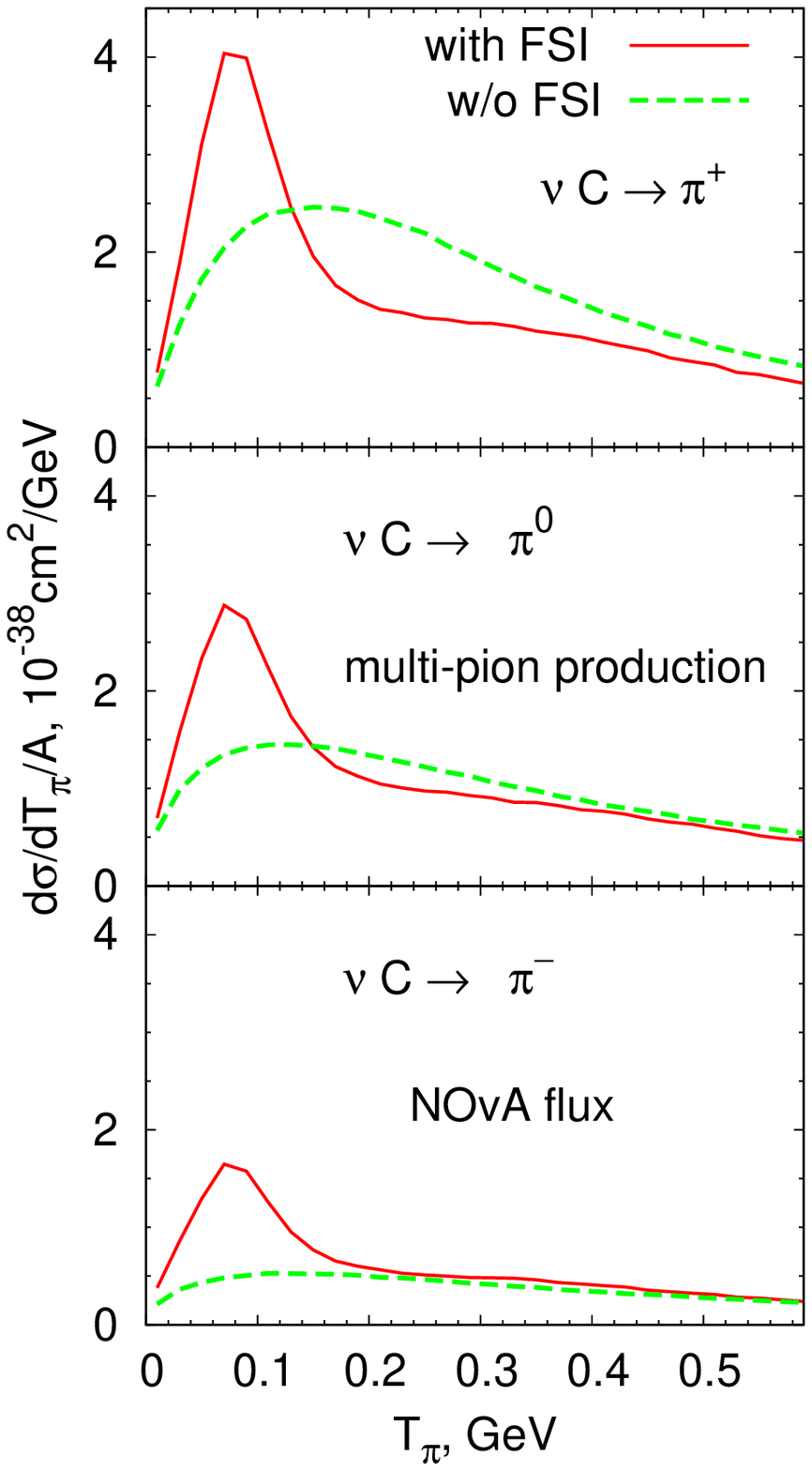}
\hfill
\caption{(Color online) Pion kinetic energy distributions  per target nucleon  for neutrino and antineutrino
scattering off iron and carbon for multi-pion production
(at least one pion of the indicated charge and any number of pions of other charges are produced).}
\label{fig:MINOS-ekin-with-wo-FSI-pion-MULTI}
\end{figure*}

Figure~\ref{fig:MINOS-ekin-with-wo-FSI-1-pion} shows the $\pi^+$, $\pi^0$
and $\pi^-$ spectra for  \onepion events
for neutrino- and antineutrino-induced reactions.
Single-pion events are
defined as those having only one pion of a given charge and no other pions in the final state.

\begin{figure*}[th]
\centering
\hfill
\includegraphics[width=0.6\textwidth]{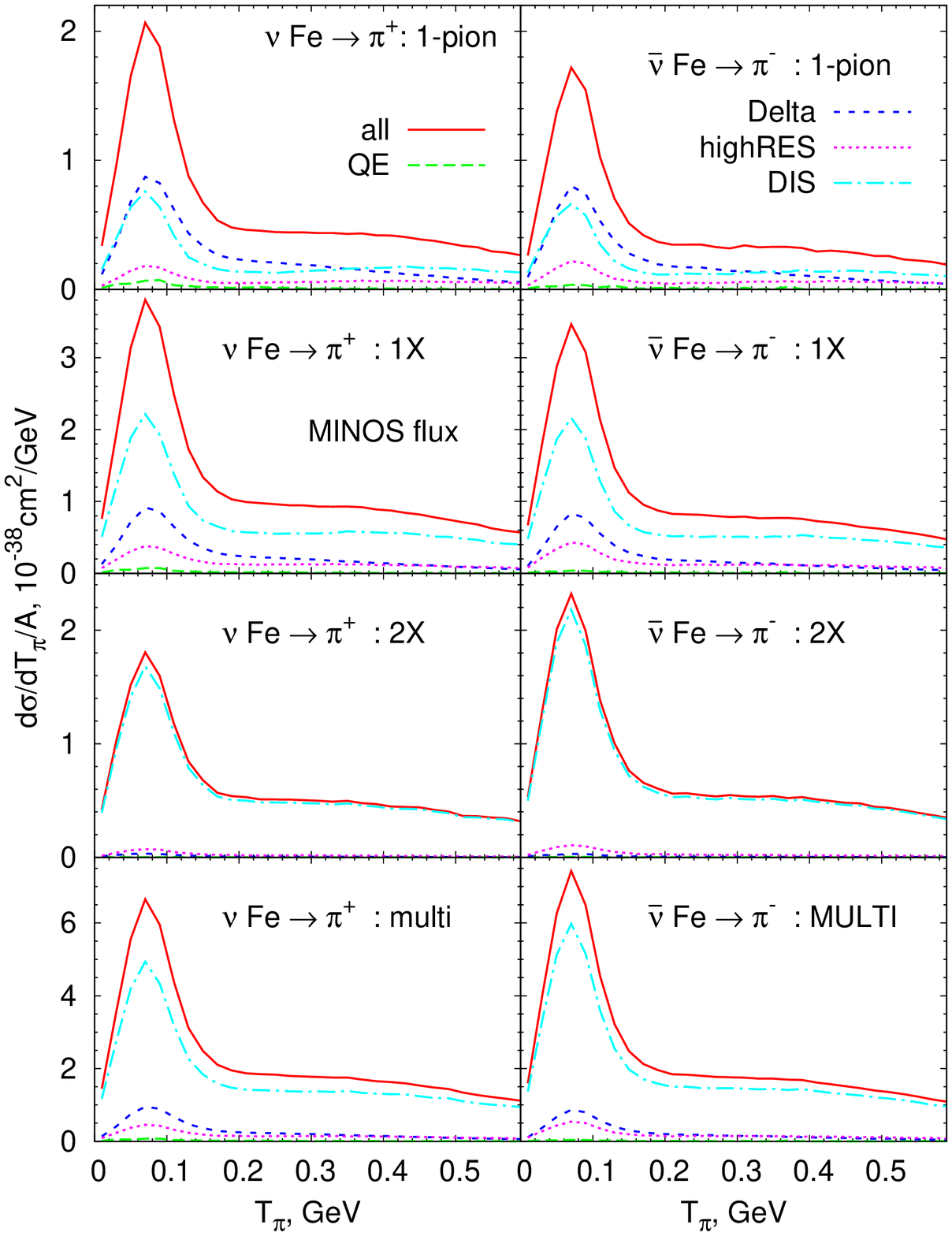}
\hfill
\includegraphics[width=0.35\textwidth]{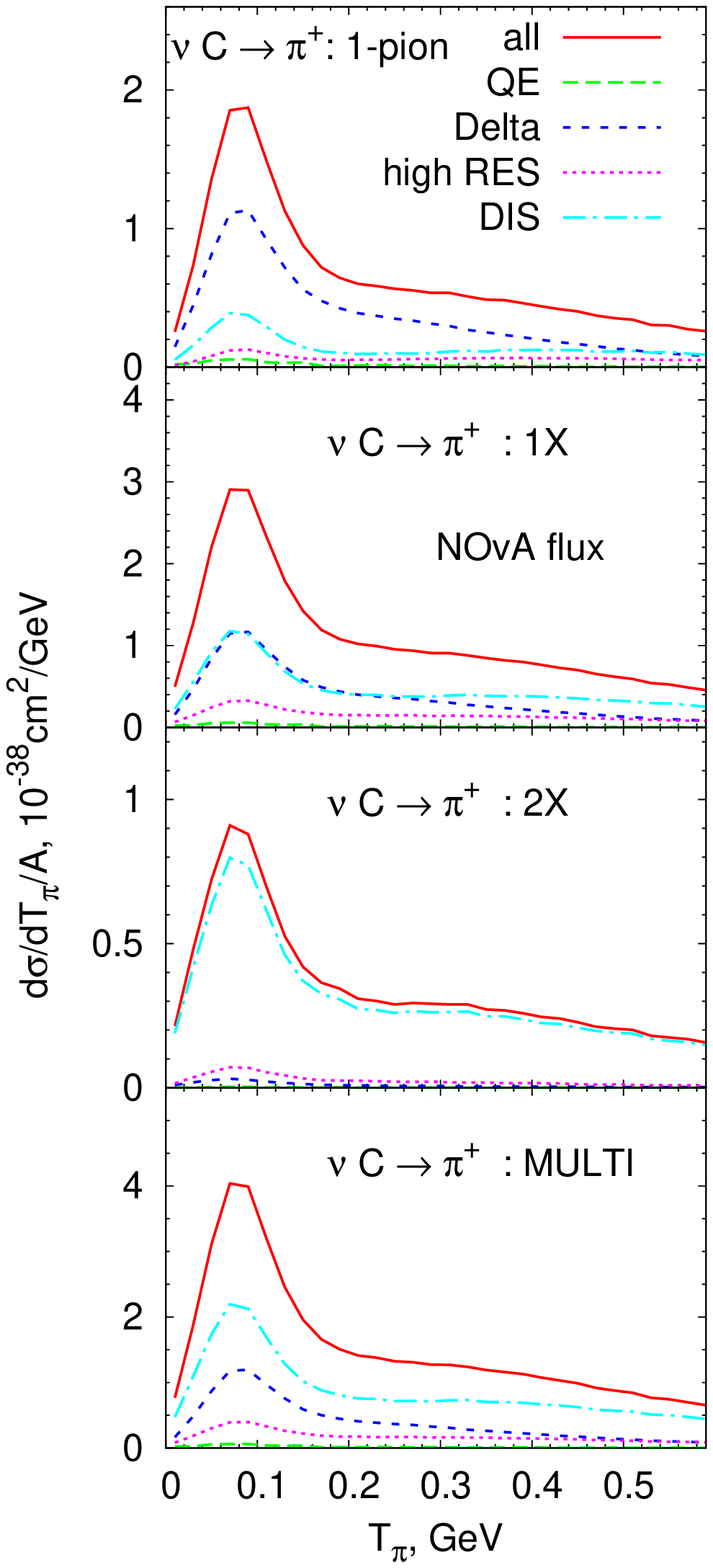}
\hfill
\caption{(Color online) Pion kinetic energy distributions  per target nucleon for $\pi^+$
production in neutrino  and $\pi^-$ production in antineutrino scattering off iron and carbon
(both are dominant channels), showing various  contribution to a given final state:
''1-pion`` only one pion of the indicated charge and no other pions are produced; ''1X`` one pion of the indicated charge and
any number of pions of other charges are produced;  ''2X`` two pion of the indicated charge and
any number of pions of other charges are produced;  ``MULTI'' at least one pion of the indicated charge and any number
of pions of other charges are produced.}
\label{fig:MINOS-ekin-variousOrigins-nu-piplus-barnu-piminus}
\end{figure*}

For the dominant channels ($\pi^+$ production in neutrino reactions and $\pi^-$ in antineutrino ones),
the FSI decrease the cross section at $T_\pi > 0.2 \GeV$. At lower neutrino energies this is mainly due to pion
absorption through $\pi N \to \Delta$ followed by $\Delta N \to NN$. At higher energies pions can also
be absorbed through $\eta$ and $\Delta$ production $\pi N\to\eta \Delta$ ,
production of higher resonances $\pi N \to R$ followed by $R N \to NN$ or $R\to \eta N$,
non-resonant pion absorption $\pi N N \to N N$, production of $\omega$ mesons
$\pi N \to \omega N$, $\phi$ mesons $\pi N \to \phi N$, and strange mesons
$\pi N \to \Sigma K, \, \Lambda K, \, K \bar{K} N$. All these channels (and more) are included in GiBUU and contribute
to pion absorption.

In addition, pion scattering in the FSI also decreases the pion energy.  Here elastic scattering as well as  DIS events of
the type $\pi N \to \mbox{multi-}\pi N$ deplete the spectra at higher energies and accumulate strength at lower energies.
Thus, an increase of the cross sections is observed at $T_\pi < 0.15 \GeV$, where the cross sections after FSI are higher than before,
and decrease above this energy.
Additionally, low-energy pions may come from the reactions such as $\eta N \to R$ followed by  $R \to \pi N$.
Altogether this leads to a significant change of the shape of the spectra. In particular, there is a strong build-up of strength around $T_\pi = 0.06 \GeV$, where the cross section after FSI is about 50\% higher than before. This is primarily due to the slowing down of pions by FSI and the low $\pi-N$ cross section in this energy region. We note that the size of this effect depends somewhat on the treatment of the collisional width of the $\Delta$ resonance \cite{Leitner:2009zz}.

Pions can also be produced in interactions of secondary nucleons. In addition, pion scattering can also involve pion charge exchange.
For neutrino-induced reactions, the  $\pi^+ n \to \pi^0 p$ scattering in the FSI is the main source of side-feeding
into the $\pi^0$ channel, leading to a noticeable increase of the $\pi^0$ cross section at low $T_\pi$.
The same effect was found for low-energy neutrino reactions \cite{Leitner:2006ww}.
Now the cross section after FSI is about 200\% higher than before. The inverse feeding is suppressed, because less
$\pi^0$ than $\pi^+$ are produced at the initial vertex. The same mechanism of side feeding
from dominant to sub-dominant channel through $\pi^- p \to \pi^0 n$
is working also for antineutrino-induced reactions.

For the least dominant channel ($\pi^-$ production in neutrino reactions and $\pi^+$ in antineutrino ones),
the FSI (in particular, side feeding into this channel)  represent the main source of the events observed.
Indeed, the $\pi^-$ yield with neutrinos  and $\pi^+$ yield with antineutrinos expected for the MINOS experiment
(cf. Fig.~\ref{fig:MINOS-ekin-with-wo-FSI-1-pion})
is, at the peak, only a little lower than that for $\pi^0$ production.

The corresponding results for the kinetic energy distribution of pions in multi-pion events, defined as consisting of at least one pion of a given
charge and any number of other pions, are shown in Fig.~\ref{fig:MINOS-ekin-with-wo-FSI-pion-MULTI}.
The cross sections here are 2-3 times higher than those of \onepion production in Fig.~\ref{fig:MINOS-ekin-with-wo-FSI-1-pion}, but the spectral distributions are very similar.

The observed pions can be produced by different mechanisms. It is,
therefore, interesting to analyze where the various final
states come from.
Fig.~\ref{fig:MINOS-ekin-variousOrigins-nu-piplus-barnu-piminus} shows the origin of the pions
(that is, the initial vertex at which the pion was produced) in the dominant channels for various final states.
It is interesting to see that even for the MINOS flux, which peaks at $3\GeV$ and has a high-energy tail, \onepion production
receives its major contribution from $\Delta$ resonance production and
its following decay. The second largest contribution comes from DIS.
This reflects the fact that even in a DIS event the final state may involve a $\Delta$ that subsequently decays into a pion.
For the other final states with more than one pion DIS dominates, but the $\Delta$ is still visible.
This reflects the fact that the pion and nucleon produced in the decay of the primary $\Delta$ undergo  FSI
which result in several pions in the final state.
The contribution from the QE vertex is very small but nonzero. In this case the outgoing pion  can
be produced only during the FSI, for example, due to the $NN\to N\Delta$ scattering followed by $\Delta\to N\pi$.

The integrated pion cross sections, averaged over the MINOS and NO$\nu$A flux distributions, are shown at the end of this section
in Table~\ref{xsec-total}. Notice here that numerical integration  of the \onepion distributions
in Fig.~\ref{fig:MINOS-ekin-with-wo-FSI-1-pion} directly leads to the values presented in the table.
For the multi-pion events this is, however, not so. Indeed, in the multi-pion distribution each pion
in a given event gives a contribution to the results in Fig.~\ref{fig:MINOS-ekin-with-wo-FSI-pion-MULTI}.
Thus, for example, an event with 3$\pi^+$ in the final state is counted 3 times in the
multi-$\pi^+$ distribution in Fig.~\ref{fig:MINOS-ekin-with-wo-FSI-pion-MULTI}.
For the integrated multi-$\pi^+$ cross section such an event is counted only once.

\subsection{Kaon kinetic energy distributions}

An advantage of the fine-grained detector used in the Miner$\nu$a experiment is
its ability to measure the energy loss in the calorimeter with high precision and, thus,
to distinguish among pions, kaons and nucleons.
Since the particle identification is based on  time-of-flight measurements,
the lower the kinetic energy, the better the separation will be; the upper boundary for distinguishing kaons
from pions is $T \approx 500\MeV$.
As we have seen, for the low-energy NuMI beam we expect that the majority of the outgoing
pions and nucleons lie below this energy region. A similar situation is expected for kaons.

At neutrino energies near the kaon production threshold and slightly above, the cross section can be described
in term of hadronic degrees of freedom and is expected to be at the level of $10^{-41}\cm^2$
for both kaon production in neutrino reactions and antikaon production in antineutrino reactions~\cite{RafiAlam:2010kf,Alam:2011dr}. In the experiments discussed here
these strangeness-changing events are clearly outnumbered by kaons produced through DIS which  dominates the kaon cross sections. Therefore, we have implemented only the strangeness conserving channels which are described by \textsc{pythia} in terms of quark and gluon degrees of freedom as $s\bar s$ pair production with the following fragmentation.

\begin{figure*}[hbt]
\centering
\hfill
\includegraphics[width=0.6\textwidth]{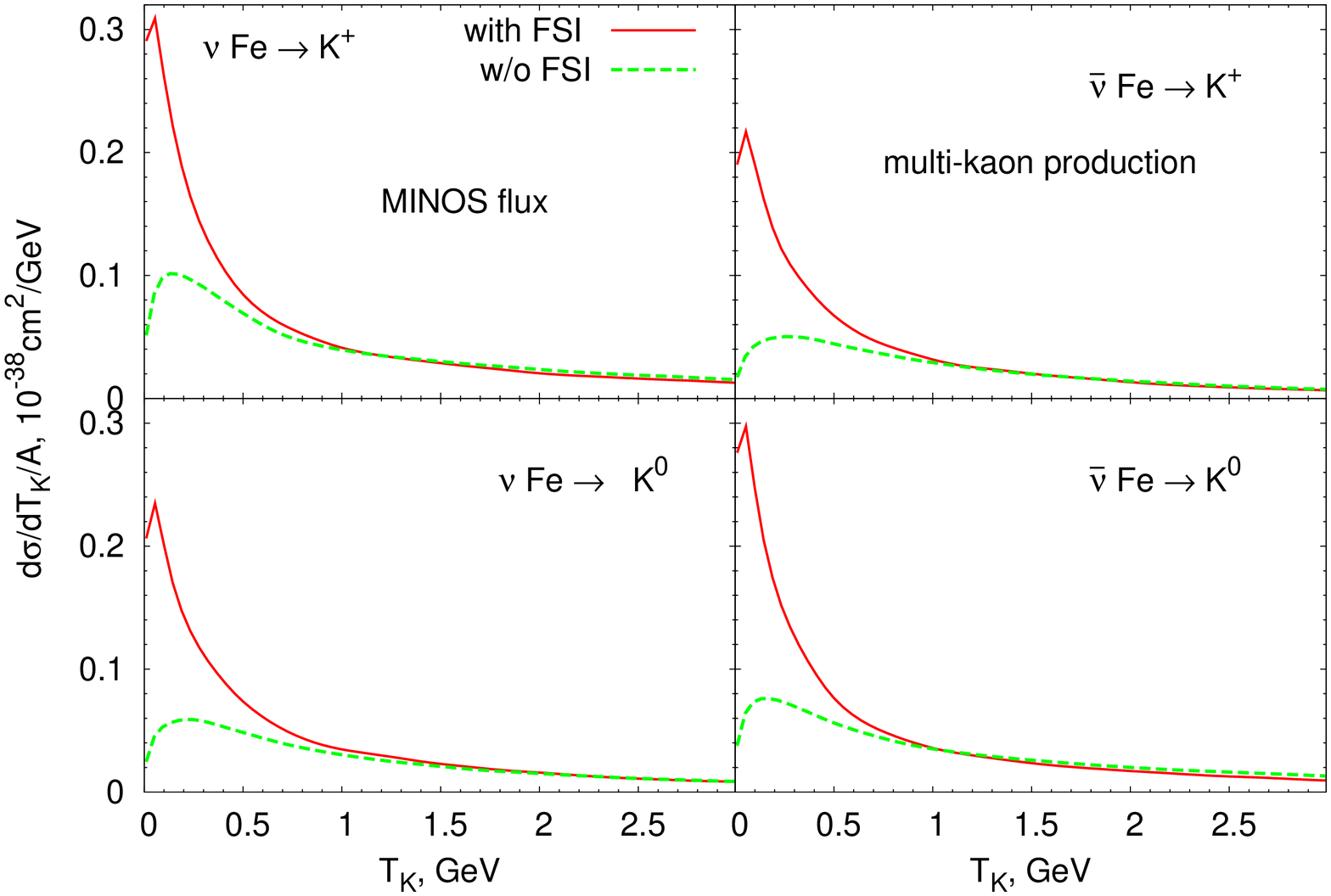}
\hfill
\includegraphics[width=0.34\textwidth]{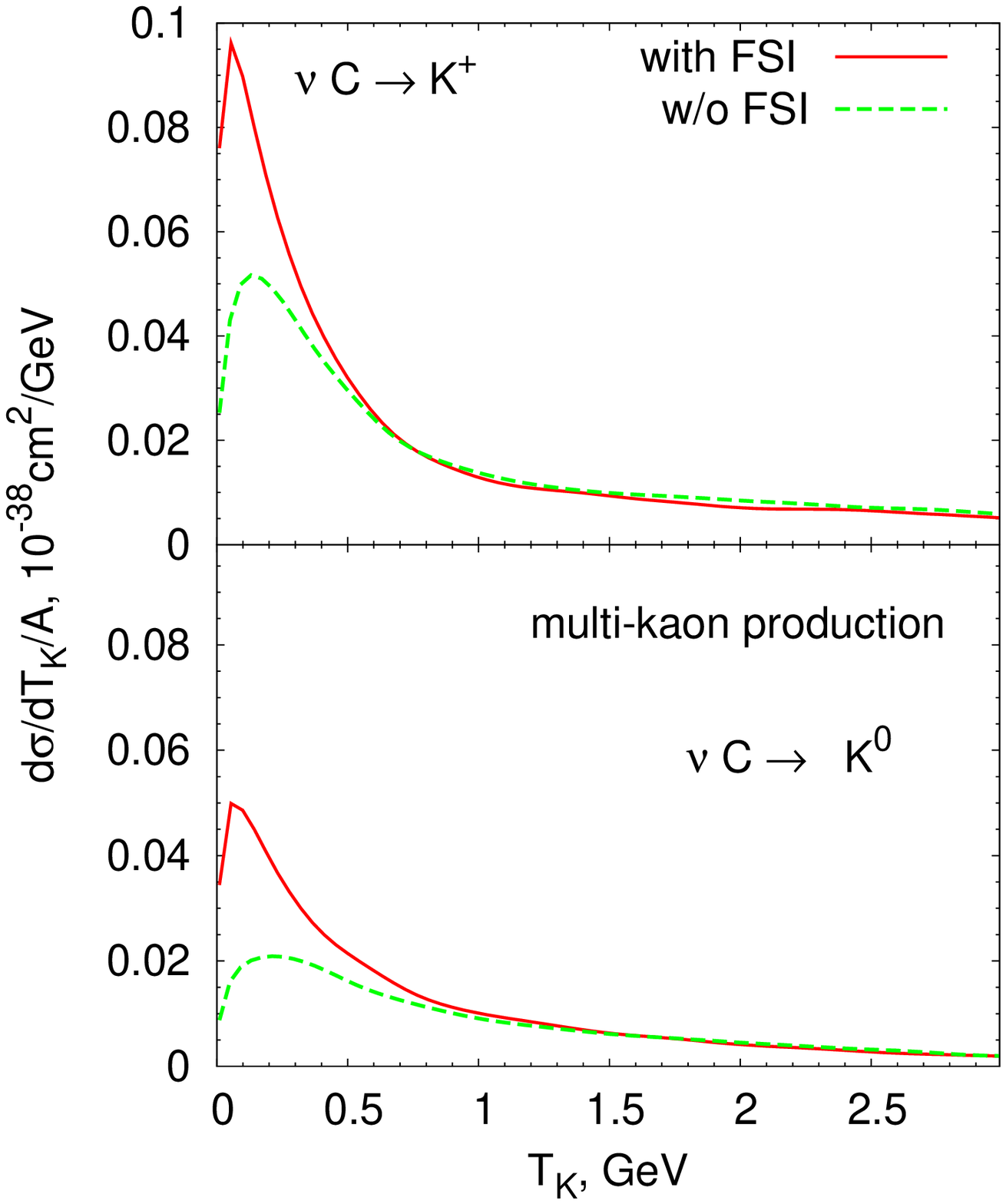}
\hfill*
\caption{(Color online) Kinetic energy distributions per target nucleon for multi-kaon
(at least one kaon of the indicated charge and any other hadrons) production
in neutrino and antineutrino scattering off iron and carbon.}
\label{fig:MINOS-ekin-with-wo-FSI-kaon-MULTI}
\end{figure*}

\begin{figure*}[hbt]
\centering
\hfill
\includegraphics[width=0.6\textwidth]{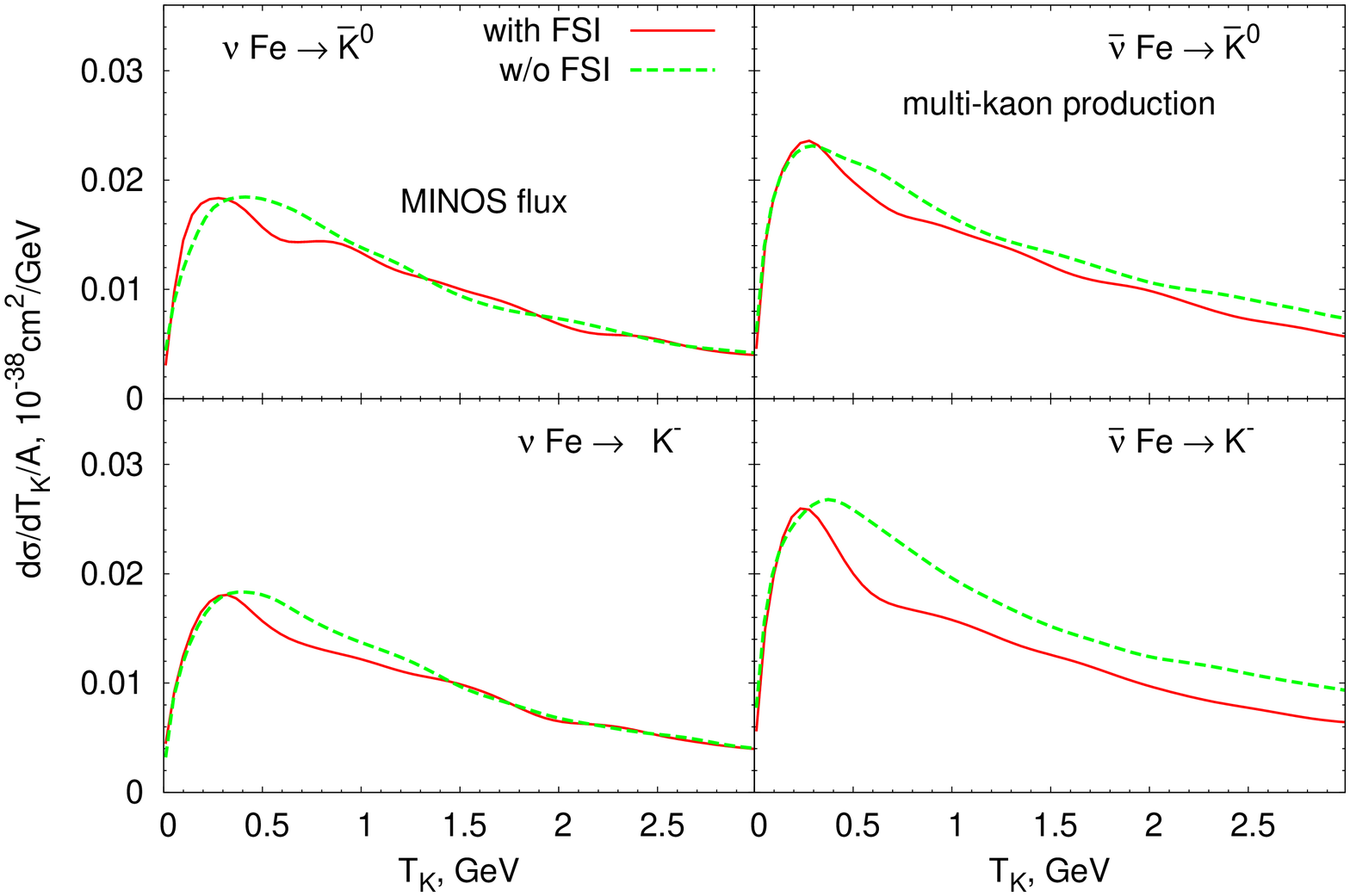}
\hfill
\includegraphics[width=0.34\textwidth]{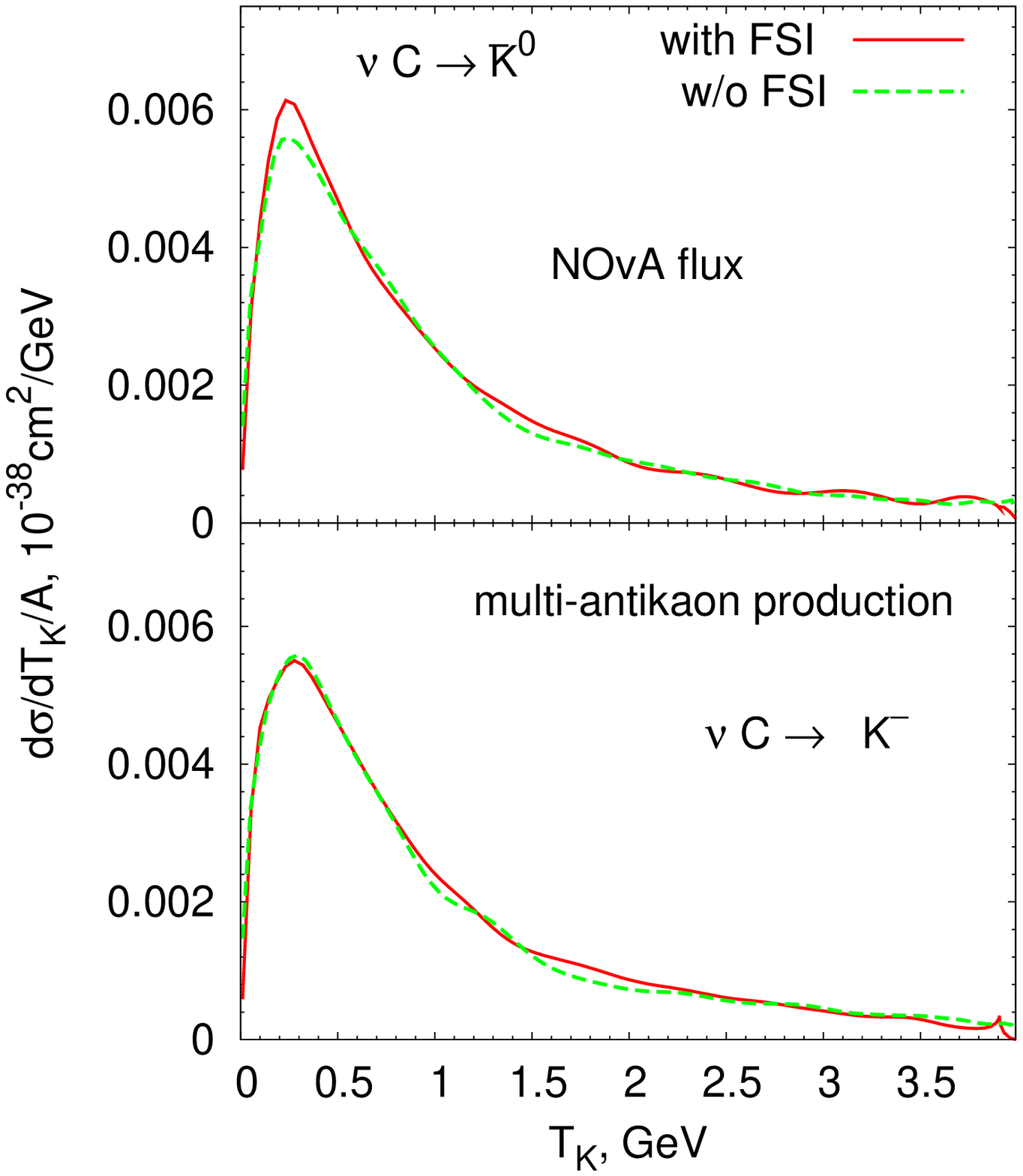}
\hfill*
\caption{(Color online) Kinetic energy distributions per target nucleon for
multi-antikaon (at least one antikaon of the indicated charge and any other hadrons)
in neutrino and antineutrino scattering off iron  and carbon.}
\label{fig:MINOS-ekin-with-wo-FSI-kaonanti-MULTI}
\end{figure*}

Before FSI, kaons can  originate from events such as $\nu + N \to K \Lambda$, $ \nu N \to K^+ K^- N$,
or more inclusive events with some pions in the final state. Due to FSI a kaon can be produced in secondary pion rescattering on nucleons
in  processes such as $\pi N\to \Sigma K, \, \Lambda K, \, K \bar{K} N $, and by nucleon rescattering via $NN\to N\Sigma K, \, N\Lambda K$.
Once produced, kaons can also rescatter elastically, $K^+ N\to K^+ N$, and inelastically, $K^+ N\to K^+ N \pi$, reducing their energy. In addition, they can undergo charge transfer, $K^+ n\to K^0 p$. Antikaons can only be produced together with kaons and they can be easily absorbed in reactions such as $K^- N \to \Lambda \pi^-$ or $K^- N \to \Sigma^0 \pi^-$ which dominate the total $K^- N$ cross section at small momenta $< \approx 0.8 \GeV$. Because these two processes proceed through anti-strange resonances these two cross sections become very large at small kaon momenta leading to a large absorption of slow antikaons.

Kaons are interesting to study because $K^+$ have a long mean free path because of strangeness conservation; they can disappear only by charge exchange into $K^0$. On the other hand, they could also be produced in secondary collisions. This makes them  sensitive to possible formation times of high-energy hadrons in the medium. In Ref.~\cite{Effenberger:1999jc} it was shown  that for  kaon-photoproduction on nuclei, which is closely related to the neutrino-induced production investigated here, the final-state interactions actually \emph{increased} the cross section for kaon production. In that same paper a significant dependence of the final kaons on the hadron formation times was also found. This was mainly due to secondary reactions of primary high-energy pions. Such pions have a very large Lorentz-boosted formation time in the nuclear target frame which suppresses their secondary interactions. As a consequence, for photon energies larger than about 3 GeV the kaon yield actually increased when the formation time for hadrons was set to zero. On the contrary, the antikaon yield decreased for zero formation time due to earlier absorption. A very similar result had been found in Refs.~\cite{Falter:2004uc,Gallmeister:2007an}. We recall from the latter calculations that the so-called leading hadrons which carry quarks from the initial reaction partners are less strongly affected by the formation times than those hadrons that are being produced from the sea \cite{Gallmeister:2007an}. Among other effects this leads to a difference in the influence of FSI on kaons and antikaons.

Fig.~\ref{fig:MINOS-ekin-with-wo-FSI-kaon-MULTI} shows the kaon kinetic energy distribution
for multi-kaon events (at least one kaon of a given charge and any number of other mesons).
Comparison of Fig.~\ref{fig:MINOS-ekin-with-wo-FSI-pion-MULTI} and Fig.~\ref{fig:MINOS-ekin-with-wo-FSI-kaon-MULTI}
shows, that the ratio of kaons to pions is expected to be at the level of $0.03-0.1$ (depending on kinetic energy).
This is very similar to the recent result of Na61/SHINE experiment on proton scattering off carbon \cite{diLuise2011}.

Since nuclei contain only positively charged and neutral nucleons, and any reaction should
conserve  strangeness, baryon number and charge, one would expect noticeably fewer antikaons than kaons in neutrino-induced reactions even before FSI.
The kinetic energy distribution for antikaons ($K^-$ and $\bar{K}^0$) is shown in Fig.~\ref{fig:MINOS-ekin-with-wo-FSI-kaonanti-MULTI}.
The corresponding cross sections are indeed 3-10 times smaller than those for kaon production in
Fig.~\ref{fig:MINOS-ekin-with-wo-FSI-kaon-MULTI}.

Interesting here is that the small antikaon cross sections do not show any
significant reabsorption, contrary to our naive expectation. For the NO$\nu$A flux the antikaon output after FSI is nearly the same as that before FSI.
A closer investigation of this result has shown that in this channel, where the cross section for the primary vertex is small, secondary production of antikaons is very important.
In other words, the antikaons seen in the detector are not the ones orginally produced in the first neutrino-nucleus 
interaction. 
With decreasing neutrino energy from high to moderate, the main source of secondary antikaons --- the  production of kaon-antikaon pairs in pion rescattering in nucleus --- becomes more and more important in comparison with the initial production.
For the MINOS neutrino flux, which peaks at about $E_\nu\approx 3\GeV$, these secondaries just compensate the absorption of antikaons initially produced at higher energies.
For the MINOS antineutrino flux the initial antikaon production is larger (because the flux has a larger part of  higher-energy component), so the absorption dominates.
For the NOVA flux, which peaks at about $E_\nu = 2\GeV$, the cross section for antikaon production is an order of magnitude lower, but the effect of secondary production is more pronounced.
For kaons both the absorption and secondary production play a less significant role.

\begin{figure}[hbt]
\centering
\includegraphics[width=0.8\columnwidth]{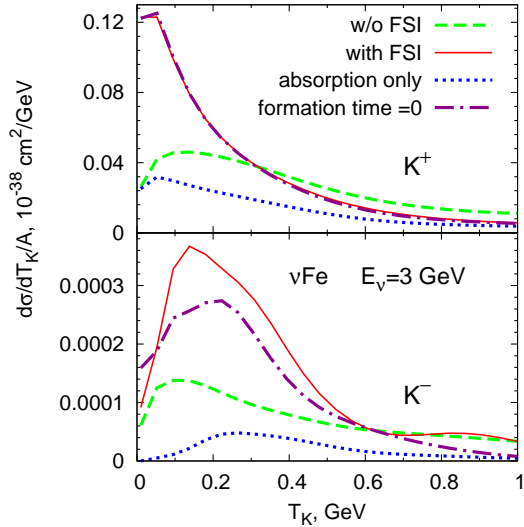}
\caption{(Color online) Kinetic energy distributions per target nucleon for
$K^+$ and $K^-$ production in neutrino scattering off iron at $E_\nu=3\GeV$.
The various curves are explained in the figure; they show the interplay between absorption and secondary production of kaons and the influence of a hadron formation time.}
\label{fig:enu3-nu-F5}
\end{figure}

The interplay between kaon absorption and production is illustrated in Fig.\ \ref{fig:enu3-nu-F5}, which shows 
the kaon and antikaon cross sections at 3 GeV, i.e.,\ the energy where the MINOS flux has its maximum. 
For $K^+$ the FSI show the expected behavior: the long-dashed line gives the spectrum of primordial kaons, i.e.,\ those produced in the initial interaction, and the short-dashed lower curve gives the spectrum as it would look like if only absorptive interactions (in this case, only charge exchange into $K^0$ is possible because of strangeness conservation) took place; this is the result a typical Glauber or optical model calculation would give. The full result  differs: Kaons with kinetic energies beyond about 0.4 GeV  are suppressed by the FSI (solid curve) compared to the initial production distribution (long-dashed). This suppression is due to rescattering with accompanying energy loss; this explains the strong rise at lower kinetic energies. These results do not change when the formation time is set to 0 because at this energy all produced hadrons (kaons and pions) are still slow enough to see the major part of the target nucleus even when they start interacting only later.
For $K^-$ the situation is considerably more complicated: Absorption alone reduces the cross section significantly, and, due to the strong rise of the total $K^-N$ cross section at small momenta the cross section goes to zero at small momenta.  The presence of secondary interactions in the FSI leads to a sizable enhancement of the yield at about $0.2 \GeV$ kinetic energy to a level well above the initial production.
In this case the influence of formation time is still small, but visible; if the $K^-$ start to interact earlier they will be more likely absorbed. Also contributing to this effect is that $K^-$ can be produced only from the sea and are thus nonleading hadrons so their interactions start with a lower cross section \cite{Gallmeister:2007an}.

The Miner$\nu$A experiment plans measurements of exclusive strangeness production. Both nuclear effects and experimental thresholds make it very difficult to distinguish truly exclusive from semi-inclusive events and to isolate the true $\nu N \to K X$ vertex. Here the same secondary production can contribute to the cross section. To extract the genuine neutrino-strangeness cross sections will require a very reliable modeling of the reaction mechanism.

To summarize this section on particle production we give the integrated cross sections averaged over the MINOS and NO$\nu$A fluxes in Table~\ref{xsec-total}.

\begin{table}[htb]
\caption{Integrated cross sections per target nucleon (in $10^{-38} \cm^2$)
for MINOS (Fe) and NO$\nu$A (C) experiments for various final states
in neutrino induced reactions}
\begin{tabular}{c@{\hspace*{7mm}}cc@{\hspace*{7mm}}cc}
\hline
            & \multicolumn{2}{c}{MINOS} & \multicolumn{2}{c}{NO$\nu$A}
\\
final state &  w/o FSI & with FSI &  w/o FSI & with FSI
\\
\hline
$1\pi^+$    &  0.76  & 0.50        	& 0.65	& 0.50
\\
$1\pi^0$    &  0.30  & 0.30		& 0.22	& 0.23
\\
$1\pi^-$    &  0.0081  & 0.14          & 0.0024	 & 0.057
\\
\hline
$1 p$       &  2.1  & 0.43             & 1.5	 & 0.62
\\
$1 n$       &  0.82  & 0.14             & 0.45	 & 0.15
\\
\hline
$1K^+$          &  0.15  & 0.18          & 0.056	& 0.061
\\
$1K^0$          &  0.095  & 0.13          & 0.027	& 0.035
\\
$1\bar{K}^0$    &  0.040  & 0.038          & 0.0068	& 0.0069
\\
$1K^-$          &  0.039  & 0.037          & 0.0065	& 0.0065
\\
\hline
$\mbox{multi }\pi^+$    &  2.2  & 1.6         & 1.3	& 1.1
\\
$\mbox{multi }\pi^0$    &  1.5  & 1.3		& 0.75	& 0.69
\\
$\mbox{multi }\pi^-$    &  0.88  & 1.0         & 0.34	& 0.42
\\
\hline
$\mbox{multi }p$       &  2.1  & 2.9            & 1.5	& 1.8
\\
$\mbox{multi }n$       &  0.83  & 2.5             & 0.45	& 1.2
\\
\hline
\end{tabular}
\label{xsec-total}
\end{table}

\section{Summary}

In this paper we have extended the GiBUU model, which was used to describe various neutrino reaction channels ---
quasielastic scattering, resonance production, and \onepion background --- to higher energies where DIS becomes relevant.
Since GiBUU provides a realistic treatment of  nuclear effects in initial- and final-state interactions and incorporates now all the relevant reaction mechanisms it is a good tool to investigate the influence of nuclear effects on observables.
For a free nucleon target the model gives results in agreement with the world average values.
For the scattering off a nuclear target, uncertainties in treating the initial-state interactions in the DIS channel lead to
uncertainty of approximately $2\%$  in the inclusive cross section, which is  compared with the recent experimental results on iron.
A study of particle knock-out events has shown that FSI, such as absorption and rescattering of the outgoing particles inside the
nucleus, are very important and lead to  a significant modification of the particle spectra.
Such spectra should be experimentally observable with the help of the fine-grained Miner$\nu$a detector.
Our results clearly indicate that, for the MINOS and NOvA fluxes, the spectra of the outgoing pions
peak below a pion kinetic energy of $100 \MeV$. For nucleons and kaons the FSI also
significantly increase the cross section at low kinetic energies. For kaons, secondary production is seen to be very important and contributes a major part of the observed cross section. It will, therefore, be extremely difficult to extract the strangeness producing $\nu N$ cross sections from experiments with nuclear targets.

\acknowledgments
This work was in its earlier stages supported by DFG.

\bibliographystyle{apsrev4-1}
\bibliography{nuclear}

\end{document}